\documentclass{aa}  

\usepackage{graphicx}
\usepackage{txfonts}
\usepackage{hyperref}	
\hypersetup{colorlinks=true,linkcolor=blue,citecolor=blue,filecolor=blue,urlcolor=blue}

\begin{document}

   \title{Determining thermal dust emission from Planck HFI data using a sparse, parametric technique}


   \author{Melis O. Irfan
          \inst{1}
          \and
          J\'er\^ome Bobin \inst{1}
          \and
          Marc-Antoine Miville-Desch\^enes \inst{1}
          \and 
          Isabelle Grenier \inst{1}
          }

   \institute{ AIM, CEA, CNRS, Universit\'e Paris-Saclay, 
       Universit\'e  Paris Diderot, Sorbonne Paris Cit\'e, 
           F-91191 Gif-sur-Yvette, France. \\
              \email{melis.irfan@cea.fr}
             }

   \date{Received XXX XX, XXXX; accepted XXX XX, XXXX}

 
  \abstract
   {The {\it{Planck}} data releases have provided the community with sub-millimetre and radio observations of the full-sky at unprecedented resolutions. 
   We make use of the {\it{Planck}} 353, 545 and 857\,GHz maps alongside the IRAS 3000\,GHz map. These maps contain information 
   on the cosmic microwave background (CMB), cosmic infrared background (CIB), extragalactic point sources and diffuse
   thermal dust emission.}    
   {We aim to determine the modified black body (MBB) model parameters of thermal dust emission in total intensity and produce all sky maps
   of pure thermal dust, having separated this Galactic component from the CMB and CIB.}
   {This separation is completed using a new, sparsity-based, parametric method which we refer to as {\texttt{premise}}. The method comprises
   of three main stages: 1) filtering of the raw data to reduce the effect of the CIB on the MBB fit. 2) fitting an MBB model to the filtered
   data across super-pixels of various sizes determined by the algorithm itself and 3) refining these super-pixel estimates into full resolution
   maps of the MBB parameters.}
   {We present our maps of MBB temperature, spectral index and optical depth at 5 arcmin resolution and compare our estimates to those of GNILC as well as 
   the two-step MBB fit presented by the {\it{Planck}} collaboration in 2013.}
   {By exploiting sparsity we avoid the need for smoothing, enabling us to produce the first full resolution MBB parameter maps from intensity measurements of thermal 
   dust emission. We consider the {\texttt{premise}} parameter estimates to be competitive with the existing state-of-the-art solutions, outperforming these methods within 
   low signal-to-noise regions as we account for the CIB without removing thermal dust emission through over-smoothing.}

   \keywords{Cosmology: Cosmic Microwave Background -- 
               ISM: dust, extinction --
                Methods: Data Analysis
               }

   \maketitle
%

\section{Introduction}

The meticulous separation of the cosmic microwave background (CMB) from astrophysical foregrounds is a pivotal step in extracting cosmological information from the {\it{Planck}} data. For intensity data several component separation techniques are currently in use; techniques such as {\texttt{SEVEM}} {\citep{sevem}}, {\texttt{NILC}} {\citep{NILC}}, {\texttt{SILC}} {\citep{SILC}},  {\texttt{SMICA}} {\citep{SMICA}} and {\texttt{L-GMCA}} {\citep{lgmca}} are specialised to recover only the CMB signal from all other emission sources. At present only {\texttt{Commander} \citep{comm}, a Bayesian parametric technique, can produce estimates of the CMB as well as the astrophysical foregrounds. Most recently, a generalised version of {\texttt{NILC}}, coined GNILC, has been applied to \it{Planck}} HFI data to recover the thermal dust emission \citep{gnilc}. This is not the case for polarisation data, where an updated version of {\texttt{Commander}, {\texttt{SMICA}} and GNILC are all capable of producing estimates for polarised thermal dust and synchrotron emission \citep{compsep18}. As our interest lies in the recovery of astrophysical foregrounds as well as the CMB and we begin our analysis using intensity data only, we initially focus on the specific problem of acquiring pure thermal dust emission maps tackled by {\texttt{Commander} and GNILC. 

Thermal dust emission is the dominant diffuse Galactic emission within {\it{Planck}} HFI data. Within the three highest HFI channels (353, 545 and 857\,GHz) the other non-negligible contributions to the total intensity are the CMB, the cosmic infrared background (CIB), extragalactic point sources and instrumental noise. Whilst the resolvable extragalactic point sources can be masked out, the unresolved extragalactic background emission that makes up the CIB requires a more global solution. The CIB makes two contributions to the total intensity measured at a particular frequency: a constant additive offset level and variations around this mean level referred to as the CIB anisotropies (CIBA). {\citet{pr2}} provide a method to determine the CIB offset level using correlations with $\rm{H_{I}}$ data; while the CIB offset level can be subtracted, the CIBA cannot. The CIBA contribution at each frequency can be approximated as Gaussian distributed noise at {\it{Planck}} resolutions ({\citet{pr2}}; \citet{gnilc}) but at scales smaller than a few arcminutes the CIBA traces structure formation at high redshifts {\citep{bethermin}}.   

Though not strictly a component separation problem, as there is only one well-defined diffuse emission source within an almost-Gaussian background, the {\it{Planck}} HFI data are an excellent testbed for techniques which aim to identify and characterise emission sources. In \citet{sparse} we presented a new parametric technique which exploits the sparse nature of astrophysical emissions within the wavelet domain yet completes the model-fit within the pixel domain. Following the work of {\citet{pr2}} and \citet{gnilc} we chose to use the modified black body (MBB) thermal dust model. \citet{compDust} present a comparison of several alternatives to the MBB model and find that none are able to fully represent the dust spectral energy distributions per unit extinction provided by \citet{pdustmod}. Therefore, without a clear successor to the MBB model across the full sky between 353 and 3000\,GHz, we also choose this simple model with its relatively small number of parameters (temperature ($T$), spectral index ($\beta$) and optical depth at 353\,GHz ($\tau_{\rm{353}}$)) required to characterise the recorded specific intensity $I_{\nu}$: 
\begin{equation}
I_{\nu} = \tau_{353} \times B(T,  \nu) \times \left(\frac{\nu}{353 \, \rm{GHz}}\right)^{\beta},
\end{equation}
where $\nu$ represents frequency in GHz and $B(T,  \nu)$ denotes the blackbody function. It should be noted that the single spectral index MBB is not the only possible model, {\citet{meis}} show that when considering frequencies both below and above 353\,GHz the use of two spectral indices (one for low frequencies, one for high) yields an improvement in the fit. The latest {\it{Planck}} HFI data processing paper \citep{datpro18}, however reveals an improvement in the calibration of the 100, 217 and 353\,GHz maps. Therefore it is possible that the need for a two-spectral index model was due to the calibration discrepancy between frequencies above and below 353\,GHz in the previous data releases. More physically motivated dust particle models are also under consideration (\citet{jones17}; \citet{gui}), as are models which account for spectral deviations from the average MBB fit along a line-of-sight using Taylor expansion \citep{chu} or, in polarisation, measurements of the 3D magnetic structure of dust clouds (\citet{tassis}; \citet{poh}). The technique presented in \citet{sparse}, herein referred to as {\texttt{premise}}, is flexible to advances in the field in that the algorithm can be updated to work with different models. Currently, though, we utilise the MBB model.

\subsection{Existing thermal dust estimates}

\citet{pr2} observe that smoothing helps reduce the impact of the CIBA on the MBB fit and so only produce their all-sky maps of dust temperature and 353\,GHz optical depth at the full 5 arcmin resolution of the {\it{Planck}} HFI data, whilst their spectral index map is at 30 arcmin. The \citet{pr2} thermal dust estimates can then be reconstructed using their MBB parameter maps. In place of global smoothing, GNILC \citep{gnilc} on the other hand, smoothes only in regions where the CIBA dominates over thermal dust emission. As they apply their filtering in the wavelet domain they can also choose to apply the greatest degree of smoothing at the angular scales which are most dominated by the CIBA. GNILC uses estimates for the CIB, CMB and instrumental noise to inform the selection of which regions to smooth. After smoothing and recomposition back into the pixel domain they can fit the MBB model, or in fact any model, to determine thermal dust model parameters. 

{\texttt{Commander} {\citep{com15}} give two 353\,GHz thermal dust estimates, both determined from an MBB fit but to different data sets. One of the dust estimates is produced at $\rm{N_{side}}$ 256 with a 1 degree FHWM from a combination of the {\it{Planck}} LFI data, WMAP {\citep{wmap9}} and Haslam {\citep{haslam}} data. The second, at $\rm{N_{side}}$ 2048 with a 7.5 arcmin FHWM, uses only the {\it{Planck}} HFI data (excluding the 100\,GHz observations) and only fits for the thermal dust spectral index, setting the thermal dust temperature at each pixel to the values determined from the low frequency fit. As {\texttt{Commander} is a Bayesian method, the thermal dust fit makes use of priors: a Gaussian distribution with a mean of 1.55 and standard deviation of 0.1 for the spectral index and a Gaussian distribution with a mean of 23\,K and standard deviation of 3\,K for temperature. 

\citet{liu} compare the thermal dust estimates produced by {\citet{pr2}}, GNILC and {\texttt{Commander} (at $\rm{N_{side}}$ 256) after all maps have been smoothed to a common resolution of two degrees. By computing the angular power spectra of the MBB parameters they identify a striking difference between the {\citet{pr2}} and GNILC spectral index power spectra at small angular scales. Although different features are visible in all three temperature and spectral index maps, \citet{liu} show that the histogram distributions for the {\citet{pr2}} and GNILC parameters are far more similar to each other than the {\texttt{Commander} distributions. 

An additional check of the MBB temperature and spectral index is their correlation. There may be a physical mechanism which produces an anti-correlation between $T$ and $\beta$; \citet{ysard} generate dust SEDs with various grain parameters and from MBB fits to these pure dust SEDs they find that the variation of grain properties can result in an anti-correlation of MBB temperature and spectral index. However, it has also been demonstrated that the presence of noise biases the MBB fits to also produce such an anti-correlation \citep{shetty}. \citet{liu} reveal both negative and positive correlations across various regions of the sky for the {\texttt{Commander} temperature and spectral indices, while the {\citet{pr2}} and GNILC temperature and spectral indices only display negative correlations. \citet{liu} note that the discrepancies between {\texttt{Commander} and both {\citet{pr2}} and GNILC are likely due to the priors imposed by {\texttt{Commander}. 

This paper is the empirical data accompaniment of {\cite{sparse}}; we use {\texttt{premise}} to determine the thermal dust MBB parameters from {\it{Planck}}, second data release (PR2), HFI data.  {\cite{sparse}} focused on this same task, demonstrating the method on simulation data where the true parameter values were known. In {\cite{sparse}} we also composed a GNILC-like algorithm of our own for comparison. Possessing the "oracle" parameter values allowed us to make to following observations: methods which rely on smoothing are at risk of smoothing out actual thermal dust information. Within the Galactic plane, {\texttt{premise}} and the GNILC-like algorithm performed comparably but in low signal-to-noise regions {\texttt{premise}} could be seen to outperform the GNILC-like algorithm. Through a combination of:
\begin{itemize}
 \item a less conservative CIBA filtering technique than GNILC,
 \item averaging over super-pixels for the MBB fit and 
 \item the removal of non-sparse wavelet coefficients,
 \end{itemize}
 {\texttt{premise}} was able to produce more accurate MBB parameters within low signal-to-noise regions than the GNILC-like algorithm. We now try this analysis on empirical {\it{Planck}} HFI data with the goal of producing all-sky MBB parameter maps: 
\begin{itemize}
\item at full resolution 
\item in less time than a pixel-by-pixel MBB fit
\item with competitive accuracy, compared to existing methods.
\end{itemize}  

In section \ref{sec:data} we present the data used for our analysis, section \ref{sec:method} gives an overview of the {\texttt{premise}} method and section \ref{sec:results} presents our results.  

\section{Data}
\label{sec:data}
In order to make comparisons between our method and existing works we perform our analysis on the {\it{Planck}} PR2 data release.

\subsection{Total intensity maps}

We determine the thermal dust MBB parameters by fitting to total intensity data at 353, 545, 857 and 3000\,GHz. The {\it{Planck}} PR2 data release provides the 353, 545 and 857\,GHz observations across the full sky at $\rm{N_{side}}$ 2048 and with a FWHM of 5 arcmin. We make use of the total intensity estimate at 3000\,GHz provided by {\citet{pr2}}. This map is a combination of the IRIS 3000\,GHz map {\citep{iris}}, made from IRAS data and the SDF {\citep{sfd}} 3000\,GHz map. This combination favours the SDF representation at scales larger than 30 arcmin and the IRIS representation at scales smaller than 30 arcmin, up to the 5 arcmin resolution of the IRIS map. We use the version of the combined IRIS and SDF map on the Planck Legacy Archive \footnotemark \footnotetext{http://pla.esac.esa.int/pla} which has had the point sources removed and inpainted.

\subsubsection{Colour corrections}

In order to account for the change of intensity across the instrumental bandpasses, colour corrections have been calculated. For the 353, 545 and 857\,GHz colour corrections of the {\it{Planck}} HFI bandpasses, the HFI bandpass responses have been used {\citep{bands}}. For the 3000\,GHz colour corrections the IRAS/SFD band responses were used \footnotemark \footnotetext{http://svo2.cab.inta-csic.es/theory/fps/}. As the colour corrections depend on the thermal dust temperature and spectral index they are calculated for each temperature and spectral index estimate within the model fitting routine. 

\subsection{Simulated CIB and instrumental noise}
\label{sec:cibnoise}
The {\texttt{premise}} method employs a similar filtering technique to that of GNILC and so requires estimates of the CIB and instrumental noise. We use the FFP8 \citep{ffp} simulations for the {\it{Planck}} HFI instrumental noise. Gaussian noise with a median level of 0.06 MJy $\rm{sr}^{-1}$ \citep{pr2} was used for the 3000\,GHz instrumental noise map. 

We also use the FFP8 estimates for the CIB at 353, 545 and 857\,GHz but we alter their mean values to match the CIB offsets measured by GNILC: CIB estimate - mean value + GNILC offset value. The GNILC CIB offset values are 0.1248, 0.3356, 0.5561 MJy/sr for 353, 545 and 857\,GHz, respectively {\citep{gnilc}}. A 3000\,GHz CIBA map of $\rm{N_{side}}$ 2048 and FWHM 5 arcmin was created using the methodology detailed in Appendix C of \citet{pr2}. We then fix the CIB offset value of this map to the GNILC calculated value of 0.1128 MJy/sr in the same way as described for the HFI frequencies. 

\subsection{CMB}
The {\texttt{premise}} method for estimating thermal dust requires for the CMB to be removed from the total intensity maps. For this we use the CMB estimate of {\citet{cmbest}}. 

\subsection{Additional data}


We make use of the \citet{green} map of interstellar reddening\footnotemark \footnotetext{http://argonaut.skymaps.info} to assess the accuracy of our optical depth at 353\,GHz estimate (section~\ref{sec:results}). \cite{green} provide E($r_{p1}$ - $z_{p1}$) measurements across 75 per cent of the sky; E($r_{p1}$ - $z_{p1}$) $\approx$ E(B-V) for $R_{V} \sim 3.1$. The map is based on stellar observations from Pan-STARRS1 \citep{panstarrs} and 2MASS \citep{twomass} data and is available as an all-sky {\texttt{HEALPix}} map at $\rm{N_{side}}$ 2048.

\section{Method}
\label{sec:method}

The motivation behind {\texttt{premise}} is the abundance of astrophysical information now available; we choose to make use of astrophysical models for the foreground components and aim to accurately determine these model parameters. {\texttt{premise}} is a sparsity driven methodology (Parameter Recovery Exploiting Model Informed Sparse Estimates) and although the data are fit in the pixel domain, the filtering and parameter refinement steps take advantage of the sparse nature of thermal dust emission within the wavelet domain. In this section we recap the {\texttt{premise}} method and detail the preprocessing steps specific to determining thermal dust emission from {\it{Planck}} HFI data. We encourage the interested reader to consult {\citet{sparse}} for the mathematical derivation of the {\texttt{premise}} algorithm.

{\texttt{premise}} has three main steps:
\begin{itemize}
\item {\bf{Filtering}} - first we filter the data to suppress the CIBA and instrumental noise. We employ the GNILC technique of using a covariance matrix formed from CIB and instrumental noise estimates but add to this method by exploiting the sparsity of thermal dust emission in the wavelet domain. As the covariance matrix is calculated for the $\rm{N_{frequency}}$ by ${\rm{N_{pixel}}}$ CIB plus instrumental noise matrix any correlations across frequencies within the noise and CIB simulations are represented. \\

\item {\bf{Super-pixel fit}} - then we avoid the computational cost of completing a pixel-by-pixel MBB fit to the filtered data by fitting to super-pixels. The unique aspect of this step being that the super-pixel areas are selected by the {\texttt{premise}} algorithm itself to ensure that the regions where the MBB parameters vary the slowest are given the largest areas to average over. An initial super-pixel area of 128X128 pixels (roughly four by four degrees) is set in order to provide the first fit to the MBB model; this "first pass" provides the algorithm with a map of reduced chi-squared values which it uses to select the final super-pixel areas. \\

\item {\bf{Parameter refinement}} - lastly we determine the optimum parameter values per original resolution pixel. Using the super-pixel, parameter maps as initial guesses we evoke a gradient descent at each pixel to minimise the least squares estimator. Once the gradient descent converges, we threshold the parameters in the wavelet domain to prevent non-sparse, noise terms from contributing to the final parameter estimates. We ensure MBB parameter maps at the original map resolution by using the empirical data, i.e. the total intensity maps containing thermal dust, CIB and instrumental noise, as the observational data in our gradient descent. The filtered maps are only used within the {\bf{Super-pixel fit}} step. 
\end{itemize}

A flow diagram summarising these three main steps is presented in Fig.~\ref{fig:meth}; as both the filtering and the super-pixel fit steps are required to obtain the initial MBB parameter estimates they are combined within the `Parameter Initialisation' block of the method flow diagram.

\begin{figure}
\centering
\includegraphics[width=0.99\linewidth]{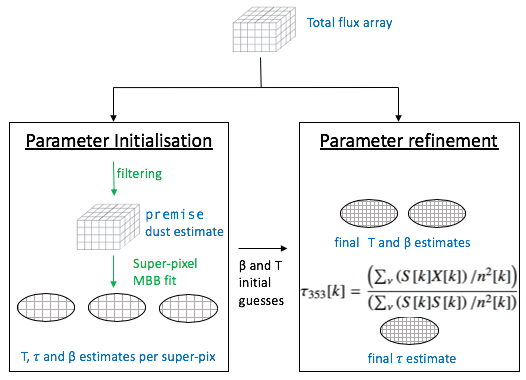}
\caption{Summary of the \texttt{premise} algorithm. Green text highlights those actions performed on the 2048X2048 pixel {\texttt{HEALPix}} faces whilst black text actions are performed on the full {\texttt{HEALPix}} sphere.}
\label{fig:meth}
\end{figure}

\subsection{Pre-processing}

We assume the CMB contribution at 3000\,GHz to be negligible and so only subtract the \citet{cmbest}} estimate of the CMB from the 353, 545 and 857\,GHz data.

The raw data are in the form of {\texttt{HEALPix}} maps. As the quadtree works in two dimensions we transform the (12X2048X2048) pixel {\texttt{HEALPix}} vector into twelve (2048, 2048) matrices and perform the filtering and super-pixel fit steps of the algorithm twelve times; once for each area of 2048X2048 pixels. The method used to convert a {\texttt{HEALPix}} vector into twelve 2D faces is described in in Appendix \ref{sec:apB}.

\subsection{Filtering}
\label{ssec:filt}

To suppress the impact of the CIBA on determining the MBB parameters, \texttt{premise} employs a modified version of the GNILC filtering technique on the CMB subtracted data. GNILC uses smoothing to suppress the CIBA within the regions of the total intensity maps, at each scale within the spherical wavelet domain, where the nuisance (instrumental noise plus CIBA) contributions dominate over the Galactic thermal dust signal. Although this technique relies upon the assumption that both the instrumental noise and CIBA can be approximated as Gaussian at each wavelet scale, the covariance matrix does take into account any slight correlations which occur over frequency.  

For {\texttt{premise}} we apply the GNILC filtering as follows for each wavelet scale ($j$):
\begin{enumerate}
 \item We define the nuisance term as N = CIB estimate + instrumental noise estimate (see section \ref{sec:cibnoise})  \\ 
 \item The raw data and nuisance estimates are divided into overlapping super-pixels of area 8X8 pixels with an overlapping ratio of $0.5$. We differ from GNILC in this respect as they choose instead to smooth the data and nuisance estimates. \\
 \item The $\rm{N_{obs}}$ by $\rm{N_{obs}}$ nuisance covariance matrix is calculated as:
\begin{equation}
{\bf{R}}_{{\rm{nus}}} = \frac{1}{\rm{Npix}} \left({\bf{N \times N^{T} }} \right)  
\end{equation}  
 \item The covariance matrices of the binned total intensity maps are calculated for each bin: 
 $ {\bf{R}}_{{\rm{tot}}} = \frac{1}{\rm{bin \, area}}{\bf{X}} \times {\bf{X}}^{T} $
 and then whitened: ${\bf{R}}_{{\rm{nus}}}^{-1/2} {\bf{R}}_{{\rm{tot}}} {\bf{R}}_{{\rm{nus}}}^{-1/2}$.\\
\item The eigenvectors of the whitened ${\bf{R}}_{{\rm{tot}}}$ are calculated for each bin and ordered. \\
\item The Marcenko-Pastur distribution is used to select eigenvalues which deviate significantly from unity. Here we again differ from GNILC, which uses the Akaike Information Criterion to select the eigenvalues which represent the signal sub-space. \\
\item The selected eigenvectors (${\bf{U}}_{s}$) give the mixing matrix (${\bf{F}} = {\bf{R}}_{{\rm{nus}}}^{1/2} {\bf{U}}_{s} $) which can be used to obtain the least-squares optimisation of thermal dust emission. We, however choose to add an additional factor to the optimisation performed by GNILC. We recapture thermal dust information lost through binning by adding a penalisation factor, to favour sparsity, into the least-squares minimisation. This penalisation factor takes the form of discrepancies between the raw data and binned dust signal ($\Delta$) which are sparse in the wavelet domain. The L2,1-norm is used to identify which discrepancies are sparse: $\sum_k \sqrt{\sum_i{{\Delta_{\nu_i}[k]^2}}}$, where $k$ is the pixel number.
\item It should be noted that a user-defined threshold value is required to identify significant wavelet coefficients, i.e. those which are high enough above the noise level to be considered signal.
\end{enumerate}

Our filtering uses two wavelet scales and a threshold cut-off value of 2.4.\\

We choose to alter the GNILC filtering as we believe that it may be too conservative: by striving to entirely remove the effects of the CIBA it may in fact be smoothing over the thermal dust signal itself. By using the Marcenko-Pastur distribution in place of the Akaike Information Criterion the \texttt{premise} filtering classifies larger regions of the full sky as signal-dominated. On top of this we then add a sparse penalisation factor to the optimisation resulting in filtered dust maps which fully capture thermal dust emission yet still contain a non-negligible CIBA contribution. This non-negligible CIBA remnant is prevented from contributing to the dust parameter estimation by the super-pixel fit.  

\subsection{Super-pixel fit}

\subsubsection{Large-scale CIB offset and point sources}

After filtering to deal with the CIBA and instrumental noise we need to remove the large-scale CIB offset before we can fit the MBB model. To enable a direct comparison between our results and that of GNILC we simply use the CIB offsets given in {\citet{gnilc}}: 0.1248, 0.3356, 0.5561 and 0.1128 MJy/sr for 353, 545, 857 and 3000\,GHz, respectively. 

The 3000\,GHz total intensity we use has already had its point sources masked and inpainted {\citep{pr2}} so we only need to mask the 353, 545 and 857\,GHz data. For this we use the 2015 {\it{Planck}} HFI point source masks.

\subsubsection{Super-pixel area selection}

We perform our MBB super-pixel fit on the filtered, large-scale CIB offset removed, point source masked data. It is worth noting that the GNILC and {\citet{pr2}} methods both apply the same MBB fitting proceeder, which includes the calibration uncertainties of the {\it{Planck}} HFI maps. Although \texttt{premise} dose not allow for calibration uncertainties within the data, we do not see a bias due to this within our results. Instead of working on the full (12X2048X2048) pixel matrices, the fit works in two dimension i.e. on each of the twelve 2048X2048 pixel areas. The procedure is as follows:
\begin{enumerate}
\item The instrumental noise + CIB covariance matrix is calculated for the full face within the wavelet domain for four wavelet scales.\\
\item The 2048X2048 face is split into patches of 128X128.\\
\item A single MBB is fit to each 128X128 patch.\\
\item The data-model residual within each patch is calculated per original pixel and transformed into the wavelet domain, again using four wavelet scales. \\
\item The reduced $\chi^{2}$ at each of the four wavelet scales is calculated for each pixel. \\
\item The data are split into super-pixels according to the reduced $\chi^{2}$; if more than ten per cent of the pixels in a 64X64 patch contain reduced $\chi^{2}$ values greater than 2 within the third wavelet scale then the patch is split into four 32X32 patches. If more than ten per cent of the pixels in a 32X32 patch contain reduced $\chi^{2}$ values greater than 2 within the second wavelet scale then the patch is split into four 16X16 patches and so on. \\
\item A lower limit of 8X8 patches is enforced and the algorithm is also prevented from cutting up the data into patch sizes which only contained masked data. \\
\item The MBB fit is then re-run on the algorithm chosen super-pixels to obtain initial parameter estimates. Fig.~\ref{fig:superpix} shows the super-pixels chosen.
\end{enumerate}

          \begin{figure}
   \resizebox{\hsize}{!}
            {\includegraphics{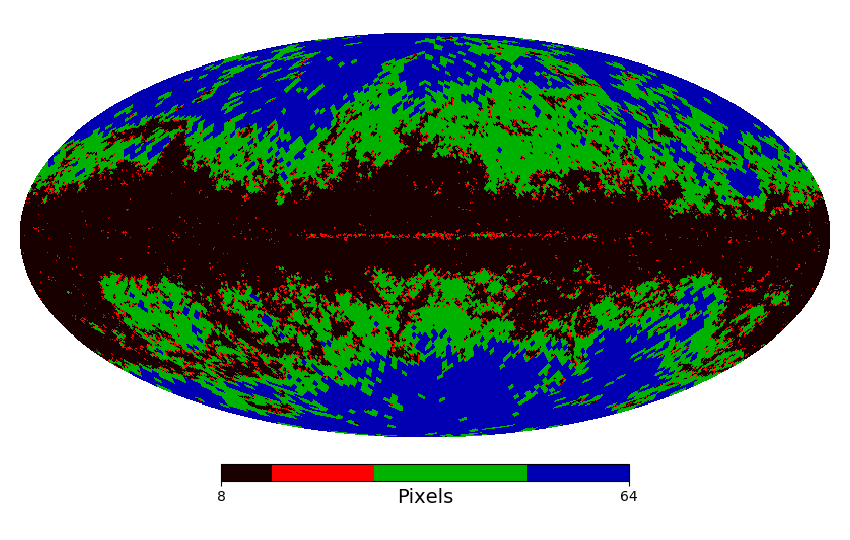}}
      \caption{All-sky map of \texttt{premise} super-pixels. Colour scale denotes pixel area: regions in black/red/green/blue are super-pixels with an area of 8/16/32/64 pixels squared.}
         \label{fig:superpix}
   \end{figure}
   
The process of dividing the total intensity maps into twelve faces, filtering these faces, selecting super-pixels and then fitting an MBB model to these super-pixels as described above is in fact repeated a total of seven times as we make use of a signal-processing technique known as `Cycle-spinning'. This technique and specifically, our application of it is described in Appendix \ref{sec:apC}. 
   
The use of super-pixels introduces an interesting feature to our method caused by the distinct spatial features of the MBB parameters. While the MBB temperature and spectral index values vary gradually across the sky, the optical depth at 353\,GHz is essentially the normalisation factor responsible for both the large and small-scale features of thermal dust. This means that for a single super-pixel, for instance a 32X32 pixel area, the MBB temperature and spectral index between 353 and 3000\,GHz may remain constant at 19\,K and 1.55 respectively whilst the optical depth traces out numerous filamentary features. Such a super-pixel fit would result in accurate estimates for temperature and spectral index, as the pixel area would be large enough to average out any residual noise and CIBA contributions, but a poorer estimate for optical depth as all variations within this area would be lost.

We deal with this trade-off between a sensitivity to the thermal dust spectral form and an accurate depiction of the normalisation factor in the parameter refinement step.  

\subsection{Refinement}

The refinement step of the \texttt{premise} algorithm is the final step and provides us with all-sky, full resolution maps of the MBB parameters. As the three MBB parameters suffer from degeneracies we shall first only discuss temperature and spectral index and then attempt to break the degeneracy, with additional data, to determine the optical depth at 353\,GHz. 

\subsubsection{Temperature and Spectral Index}

   \begin{figure*}
      \centering
            {\includegraphics[width=0.71\linewidth]{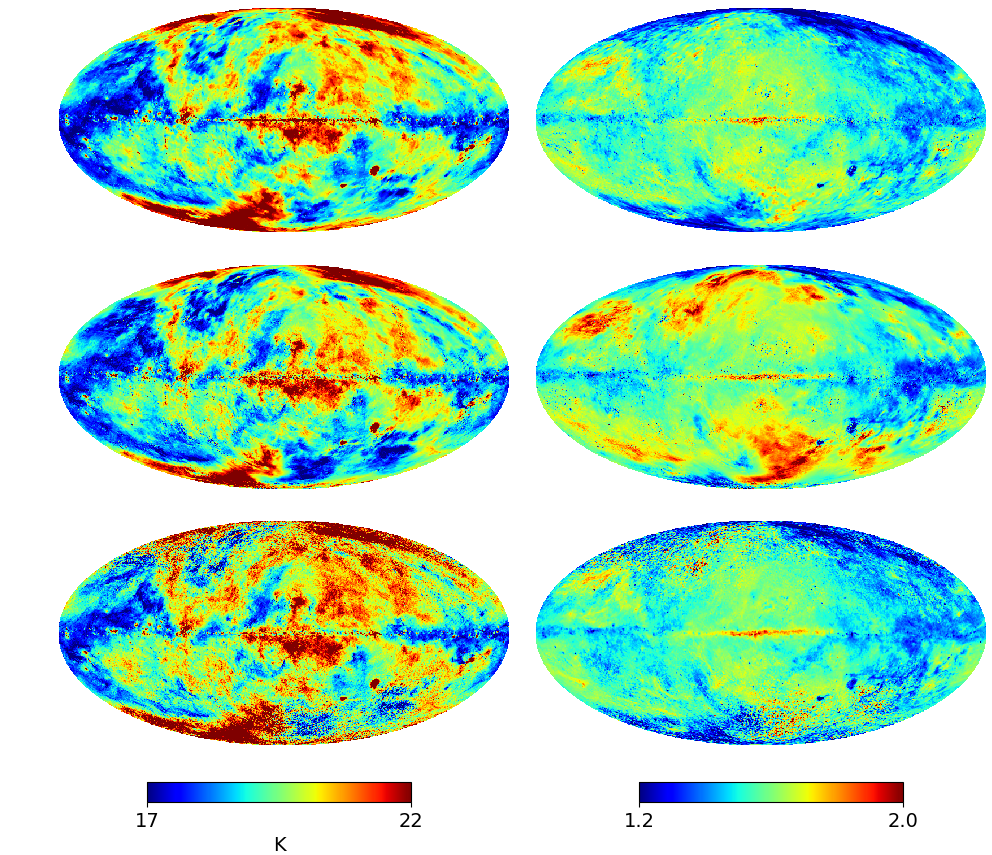}}
      \caption{Thermal dust MBB temperatures ({\it{left}}) and spectral indices ({\it{right}}) as calculated by \texttt{premise} ({\it{top}}), GNILC ({\it{middle}}) and 2013 ({\it{bottom}}).}
         \label{fig:temps}
   \end{figure*}
   
      \begin{figure*}
      \centering
            {\includegraphics[width=0.71\linewidth]{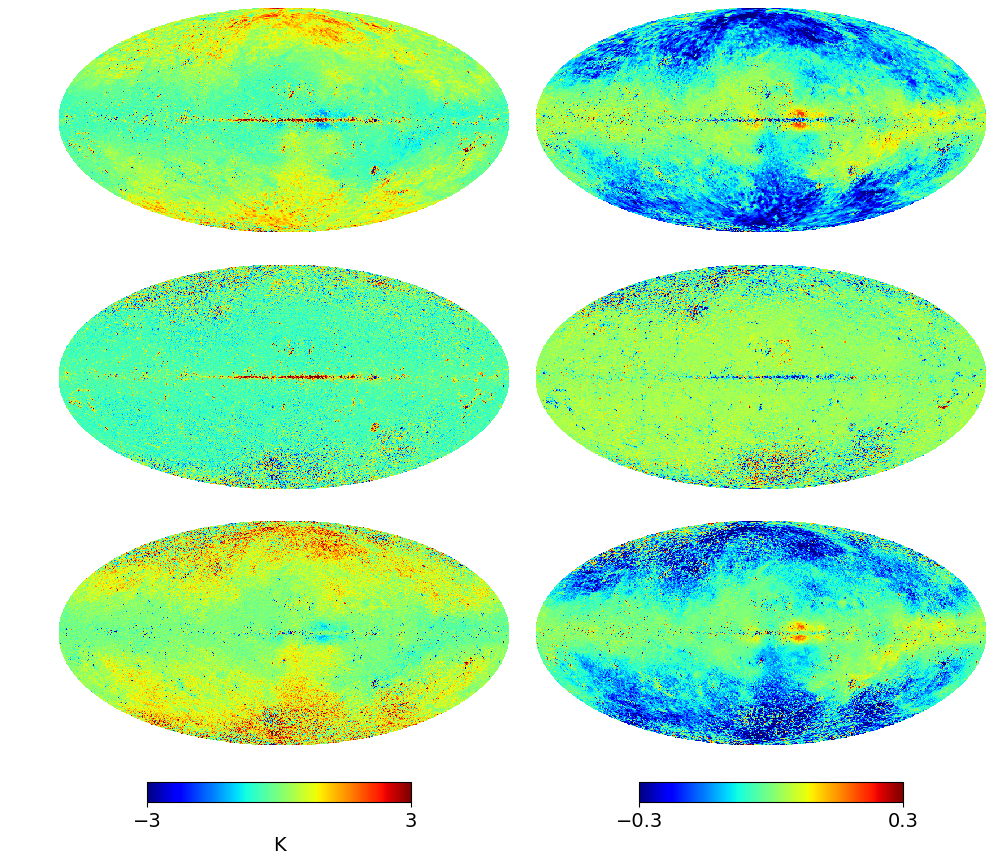}}
      \caption{Differences in thermal dust MBB temperatures ({\it{left}}) and spectral indices ({\it{right}}). \texttt{premise} minus GNILC ({\it{top}}), \texttt{premise} minus 2013 ({\it{middle}}) and 2013 minus GNILC ({\it{bottom}}).}
         \label{fig:tempsdiff}
   \end{figure*}

The super-pixel fit produces initial estimates of the MBB temperature and spectral index. As these initial estimates are well informed we can use them to seed a gradient descent algorithm to determine the optimum temperature and spectral index per pixel. First we calculate an initial guess for optical depth at 353\,GHz by minimising the reduced $\chi^{2}$ at each 5 arcmin pixel ($k$) between the total intensity (thermal dust plus CIB plus instrumental noise) data and our spectral model of thermal dust emission ($S$). Our spectral model is formed using the quadtree estimates for temperature and spectral index produced by the super-pixel fit of the algorithm. We weight the data according to our CIBA plus instrumental noise estimates (n): 
\begin{eqnarray}
\label{taueq}
\tau_{353}^{0}[k] = \frac{\left( \sum_{\nu} \left ( S[k] X[k] \right) / n^{2}[k] \right)}{\left(  \sum_{\nu} \left ( S[k] S[k] \right) / n^{2}[k] \right) },
\end{eqnarray}
where $X[k]$ is a vector containing the total intensity measurement at pixel $k$ at each frequency and  
\begin{eqnarray}
S[k] = B(T[k],  \nu) \times \left(\frac{\nu}{353 \, \rm{GHz}}\right)^{\beta[k]} \times 10^{20} \times CC^{-1}[k], 
\end{eqnarray}
where $CC^{-1}$ are the inverse colour corrections for the four bandpasses and the factor of $10^{20}$ is the conversion factor from power to MJy. We can also calculate a noise estimate as follows:
\begin{eqnarray}
\label{taueqnoise}
\delta \tau_{353}[k] = \frac{\left( \sum_{\nu} \left ( S[k] n[k] \right) / n^{2}[k] \right)}{\left(  \sum_{\nu} \left ( S[k] S[k] \right) / n^{2}[k] \right) },
\end{eqnarray}
We then produce our initial $\tau_{353}$ estimate by soft thresholding the map in the wavelet domain, setting the threshold at 2.4 times the noise estimate ($\delta \tau_{353}$). 

We can then construct a thermal dust model at each frequency ($\nu$) as $\tau_{353}^{0} \times B(T^{0}, \nu) \times (\nu/353)^{\beta^{0}}$. The gradient descent is performed iteratively, per pixel, for both temperature and spectral index as follows:
\begin{equation}
\beta/T^{(t+1)}[k]  =  \beta/T^{(t)}[k] +  \alpha \left(\frac{\partial  X[k]}{\partial \beta/T} \right) ^{T} \, (X[k] - {\rm{Model}}[k]).
\end{equation}
The gradient path length ($\alpha$) is computed analytically at each iteration; it is the minimum of the inverse Hessian matrix of the data-model residual (with respect to $\beta$ and $T$). The gradient descent is terminated upon convergence, which we define as when two iterations only differ in their parameter estimation by less than $10^{-6}$.

After the gradient descent step we now have two all-sky maps, one of temperature and one of spectral index. These maps are at the full {\it{Planck}} resolutions of 5 arcmin as the gradient descent step uses the total intensity (thermal dust plus CIB plus instrumental noise) data, not the filtered data from section \ref{ssec:filt}. Using the spherical wavelet transform we soft threshold these maps in the wavelet domain (leaving the coarse scale untouched) to remove the effect of any non-sparse elements in the total intensity maps. The threshold values are calculated at each of the five wavelet scales as twice the mean absolute deviation (MAD) of the temperature/spectral index. As variance cannot be calculated per pixel we divide the data into patches of 32X32 pixels for the MAD calculation. Thresholding removes non-sparse information from the wavelet scales and so if, for a particular region and angular scale, the signal is completely subdominant to the noise then we loose the signal for that region at that resolution. We also exclude information from the first wavelet scale as this corresponds to individual pixel variations, so changes in temperature and spectral index across scales smaller than the {\it{Planck}} beam FWHM. These spurious variations are produced by the inclusion of the CIBA in the total intensity data used for the gradient descent. 

We apply a total point source mask, made from the multiplication of the 353, 545 and 857\,GHz {\it{Planck}} HFI masks, to the residual matrix ($(X - {\rm{Model}})$) in order to prevent the point sources from contributing to the final temperature and spectral index estimates. Therefore, temperature and spectral index values within the total point source mask remain at the initial, super-pixel estimate. The process of gradient descent followed by thresholding within the wavelet domain is then repeated to ensure a robust solution. 

\subsubsection{Optical depth at 353\,GHz}
The MBB parameters are heavily degenerate when fit in the presence of CIBA and noise. Both {\citet{pr2}} and {\cite{gnilc}} try to combat this degeneracy through smoothing. The {\texttt{premise}} super-pixel MBB fit is specialised to select super-pixels sizes based on the spatial variation of the dust temperature and spectral index, as of such the same method is less responsive to the overall normalisation factor (the optical depth).

To break the parameter degeneracies we return to the total intensity (thermal dust plus CIB plus instrumental noise) data for additional information. We calculate our optical depth estimate at 5 arcmin using Eq.~\eqref{taueq} but this time our spectral model is formed using the 5 arcmin estimates for temperature and spectral index produced by the refinement step of the algorithm. As before, we can calculate a noise estimate using Eq.~\eqref{taueqnoise} and produce our final $\tau_{353}$ estimate by soft thresholding this map in the wavelet domain, setting the threshold at 2.4 times the noise estimate ($\delta \tau_{353}$). 

      \begin{figure}
      \centering
            {\includegraphics[width=0.7\linewidth]{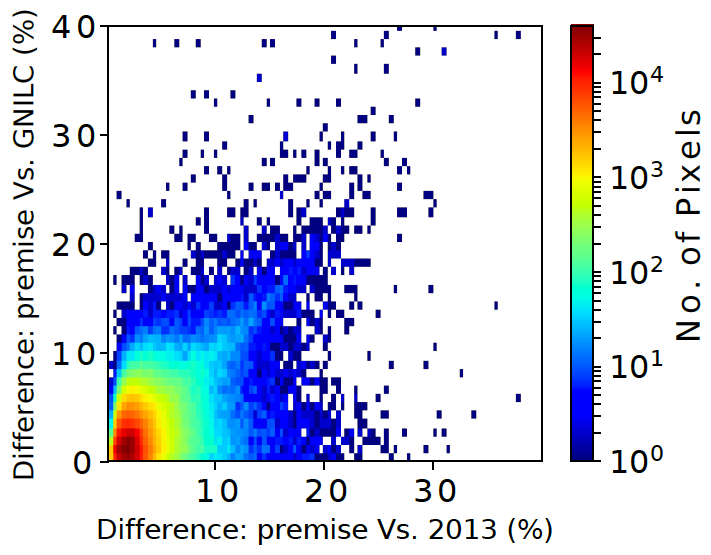}} \\
            {\includegraphics[width=0.7\linewidth]{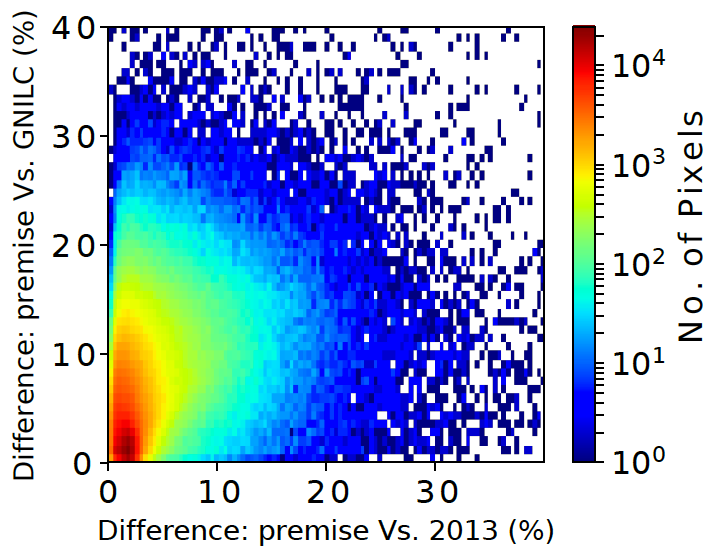}} 
      \caption{Percentage difference between the \texttt{premise} and GNILC temperature ({\it{top}})/spectral index ({\it{bottom}}) values against the percentage difference between the \texttt{premise} and 2013 temperature/spectral index values.}
         \label{fig:tempspecdiff}
   \end{figure}

\section{Results}

We aim to assess our thermal dust MBB parameters through comparisons with existing thermal dust templates and external data sets. Within this section we shall refer to the results obtained by {\citet{pr2}} as `2013'; whilst the method dates from this time the results themselves have been reproduced here using PR2 data and it is these updated results which we use in this work.

\label{sec:results}
\subsection{Temperature and Spectral Index: spatial distributions}

Fig.~\ref{fig:temps} shows the all-sky MBB temperature and spectral index for thermal dust emission as calculated by \texttt{premise}, 2013 and GNILC. The differences between the parameter estimates are given in Fig.~\ref{fig:tempsdiff}. Visually, for both the temperature and spectral index maps, the \texttt{premise} and 2013 maps look the most similar, differing from the GNILC  temperature and spectral index maps noticeably at high latitudes. The globular features, clearly seen in any of the difference maps which include the GNILC spectral index estimate, are discussed in detail in Section ~\ref{sec:dustmaps}.

In Fig.~\ref{fig:tempspecdiff} we plot the percentage difference between the \texttt{premise} and GNILC temperature/spectral index values against the percentage difference between the \texttt{premise} and 2013 temperature/spectral index values. For Fig.~\ref{fig:tempspecdiff} we calculate the percentage difference for the $\rm{N_{side}}$ 2048 maps but then downgrade the results to $\rm{N_{side}}$ 256 for the plot as we aim to highlight trends over regions of the sky as opposed to individual anomalous pixels. A `spike' of linear correlation is seen in the temperature (top) plot of Fig.~\ref{fig:tempspecdiff}, this corresponds to a thin strip within the Galactic plane where the abundance of point sources prevents \texttt{premise} from providing accurate temperature estimates. Masking of this region removes the spike from the temperature density plot. Otherwise, there are no other regions where GNILC and 2013 consistently disagree with \texttt{premise} over the parameter values; implying that the discrepancies in parameter values are mainly due to which methodology is used. 

          \begin{figure}
          \centering
              {\includegraphics[width=0.99\linewidth]{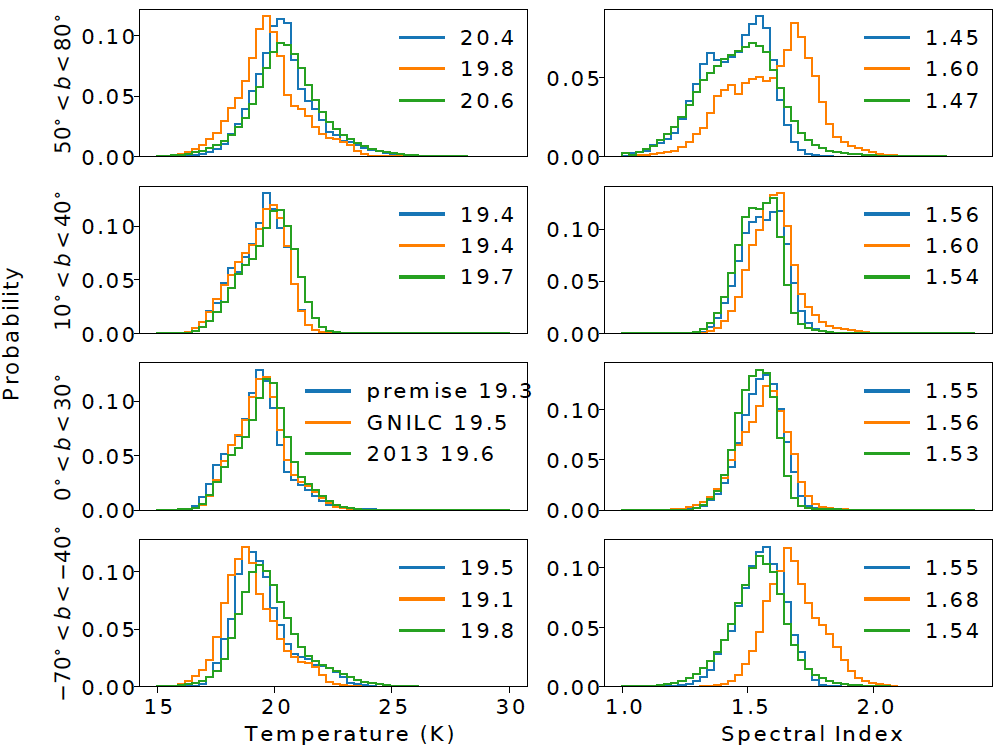}}
      \caption{Histogram comparison between \texttt{premise} (blue), 2013 (green) and GNILC (red) temperatures ({\it{left}}) and spectral indices ({\it{right}}) for different strips in Galactic latitude.}
         \label{fig:histSpec}
   \end{figure}
   
The mean spectral index and temperature values across the full sky are 19.5\,K and 1.54 respectively for \texttt{premise}, 19.4\,K and 1.60 respectively for GNILC and 19.8\,K and 1.53 for 2013. To further probe the discrepancies in parameters estimates, Fig.~\ref{fig:histSpec} shows the distributions of the temperatures and spectral indices within various strips of Galactic latitude as calculated by \texttt{premise}, 2013 and GNILC. The mean parameter estimates within each latitude strip are noted on each histogram. At high Galactic latitudes ($|b|> 40^{\circ}$) the GNILC temperatures estimates are lower than those of 2013 or \texttt{premise}, whilst close to the Galactic plane the three temperature estimates are close to indistinguishable. The three spectral index estimates are also fairly similar close to the Galactic plane, varying by 0.06 within $ 10^{\circ}< b < 40^{\circ} $ and 0.03 within $0^{\circ}< b < 30^{\circ}$. The largest differences in spectral index estimates are seen at high latitudes where the GNILC estimates favour higher values than those of 2013 or \texttt{premise}}}. Once again, the three sets of parameter estimates differ most notably within the regions where the CIBA and instrumental noise contribute most significantly to the total intensity.

              \begin{figure}
   \centering
            {\includegraphics[width=0.99\linewidth]{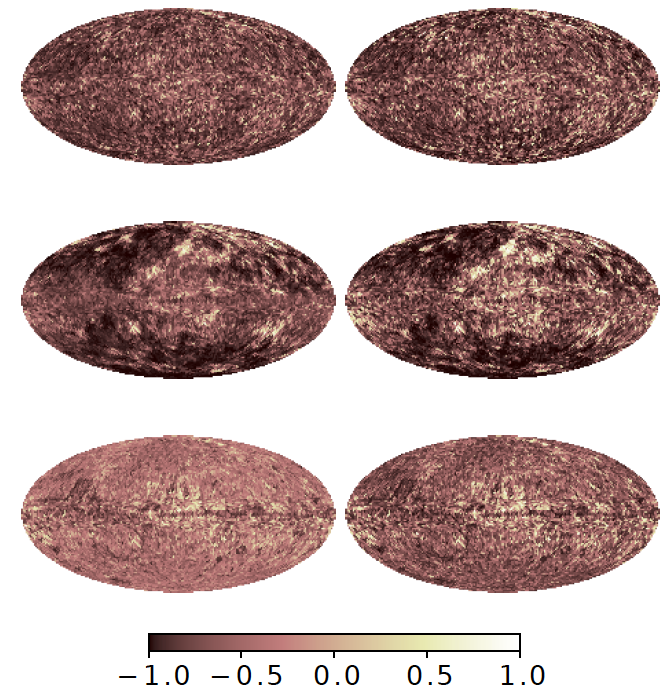}}\\
      \caption{Temperature and spectral index correlation maps for \texttt{premise} parameters {\it{(top}}), GNILC parameters {\it{(middle}}) and 2013 parameters ({\it{bottom}}). The colour scale represents the correlation coefficient values. The left-hand column shows the parameters at their original resolutions whilst the right-hand correlation maps are made from parameters smoothed to 30 arcmin.}
         \label{fig:corrsfull}
   \end{figure}

\subsection{Temperature and Spectral Index: anti-correlations}

A possible check of MBB temperature and spectral index is their correlation. While it may be possible for there to be a physical mechanism which produces an anti-correlation between $T$ and $\beta$, it is known that the presence of noise and the CIBA biases the MBB fits to also produce such an anti-correlation \citep{shetty}. Additionally, it is worth noting that the MBB is an approximated model of  thermal dust emission.  The very action of imposing an MBB fit onto a physical model not perfectly described by an MBB fit also introduces such anti-correlations. 

Fig.~\ref{fig:corrsfull} shows correlation coefficient maps made by dividing the full sky into patches of 64X64 pixels, giving just under two degrees of resolution. Each patch is coloured according to the correlation coefficient determined for the 4096 pixels within the area. The 2013 $\beta$ estimates are presented at a resolution of 30 arcmin across the full sky, while their temperature estimates are at 5 arcmin. The \texttt{premise} temperature and spectral index estimates are at 5 arcmin resolution across the full sky and the GNILC temperature and spectral index estimates vary from 5 arcmin close to the Galactic plane to 22 arcmin within the lowest signal-to-noise regions. The correlation maps in the left-hand column of Fig.~\ref{fig:corrsfull} are made from the parameters at their original resolutions whereas the maps in the right-hand column are made from parameter maps smoothed to 30 arcmin.

In the left-hand column correlation maps larger patches of correlated and anti-correlated pixels appear at high latitudes (low signal-to-noise regions) for GNILC than for \texttt{premise} while the 2013 temperature and spectral index values can be seen to be barely correlated. Interestingly, all three plots show strong positive correlations around the Galactic centre which are unlikely to be caused by noise or the CIBA as the Galactic plane has the highest signal-to-noise ratio within the full sky. The same is seen for the right-hand maps; the main difference observed is the increase in magnitude of existing positive/negative correlations. This is because smoothing to 30 arcmin averages out some of the noise (as well as some of the signal) within the parameter maps.

In Fig.~\ref{fig:corrplot} we select several interesting regions, such as one of the areas displaying positive correlations and examine the $T$-$\beta$ anti-correlation in the form of contour plots. The regions are circles of radius two degrees, centred at $(l,b) = (60^{\circ}, 50^{\circ}), (0^{\circ}, 3^{\circ}), (200^{\circ}, -5^{\circ})$ and $(150^{\circ}, 25^{\circ})$. The pixels are binned and the one sigma contour levels are plotted. The correlation coefficients for the pixels within each regions are given on each plot.
   
      \begin{figure}
   \centering
            {\includegraphics[width=0.99\linewidth]{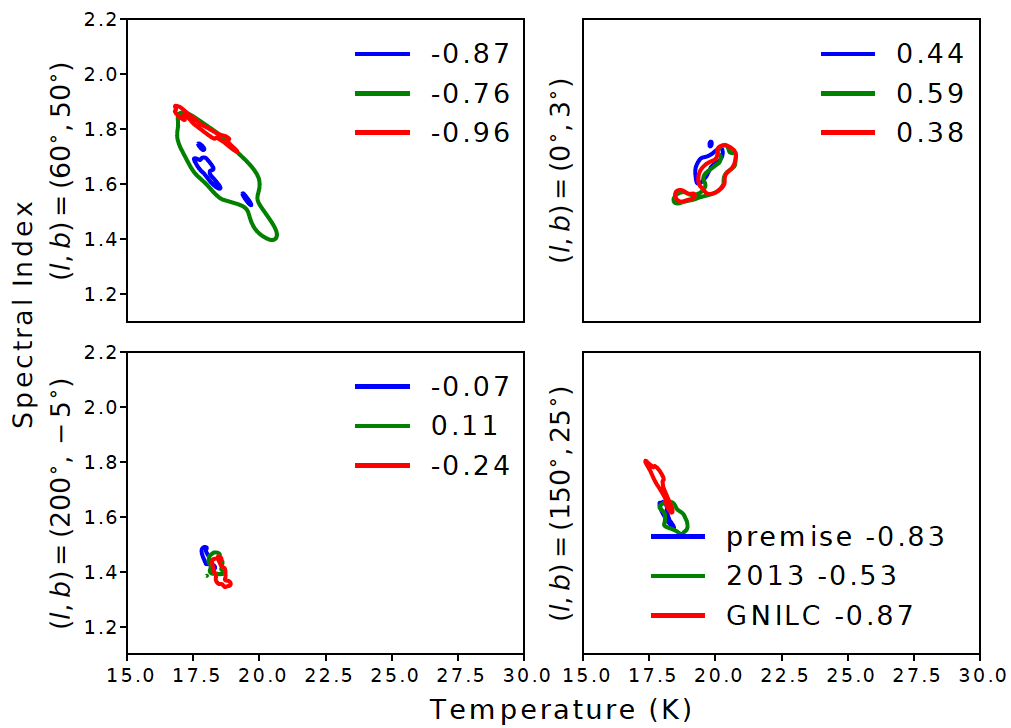}}\\
      \caption{Contour plots of temperature and spectral index for \texttt{premise}, GNILC and 2013. The one sigma level contour is shown for a two degree radius circle of pixels centred at various coordinates.}
         \label{fig:corrplot}
   \end{figure}

      \begin{figure*}
   \centering
            {\includegraphics[width=0.9\linewidth]{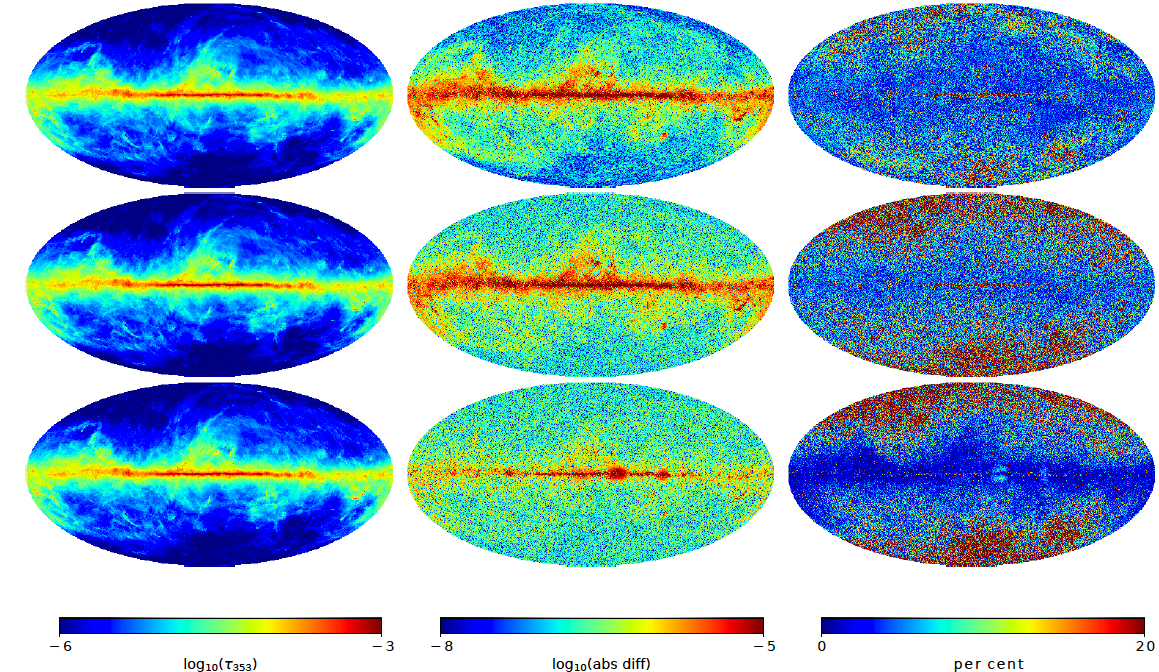}}\\
      \caption{ {\it{First column}}: The optical depth at 353\;GHz ({\it{left}}) as calculated by \texttt{premise} ({\it{top}}), GNILC ({\it{middle row}}) and 2013 ({\it{bottom}}). {\it{Second column}}: The differences between the optical depth at 353\;GHz estimates, \texttt{premise} minus GNILC ({\it{top}}), \texttt{premise} minus 2013 ({\it{middle row}}) and 2013 minus GNILC ({\it{bottom}}). {\it{Third column}}: The percentage differences between the optical depth at 353\;GHz estimates.}
         \label{fig:alltau}
   \end{figure*}

The first region ($(l,b) = (60^{\circ}, 50^{\circ})$) is one of low signal-to-noise. Here the 2013 $T$-$\beta$ values occupy a larger parameter space indicating noisier measurements which, as the correlation coefficient is inversely proportional to the standard deviation of the $T$-$\beta$ values, results in lower coefficient magnitudes. The GNILC parameter sets display the largest anti-correlation. The second region ($(l,b) = (0^{\circ}, 3^{\circ})$) was selected as it is one of the high signal-to-noise regions were all three parameter sets display positive correlation. The third region ($(l,b) = (200^{\circ}, -5^{\circ})$)  represents a typical high signal-to-noise region; all three parameter sets display negligible anti-correlation though the GNILC set still posses the largest negative correlation coefficient. The fourth region ($(l,b) = (150^{\circ}, 25^{\circ})$) was selected as all three parameter sets were seen to show strong negative correlations despite the high signal-to-noise ratio in this region. For regions like region 2 and region 4, where the signal-to-noise ratio is high yet significant positive/negative parameter correlations are seen for all three methods, it is possible that the thermal dust emission itself is less successfully categorised by the MBB model.

\begin{table}
   \caption{The properties of several circular regions of radius two degrees within parameter space. The columns show the central coordinates of the regions, the correlation coefficient between temperature and spectral index, the standard deviation in temperature and the standard deviation in spectral index. For each region the \texttt{premise} correlation coefficient and standard deviations are listed first, then the GNILC values, then the 2013 values.}
   \label{table:2}
   \centering
   \begin{tabular}{c c c c}
   \hline \hline
   ${\bf{(l,b)}}$ (${\bf{^{\circ}}}$) & ${\bf{r}}$ & ${\bf{\sigma T}}$ {\bf{(K)}} & ${\bf{\sigma \beta}}$  \\
   \hline
    & \multicolumn{2}{c}{{\bf{Low signal-to-noise}}} \\
   (60, 50) & -0.87 & 0.69 & 0.07 \\
     & -0.96 & 0.71 & 0.05 \\ 
    & -0.76 & 1.19 & 0.15 \\
   (60, -60) & -0.52 & 0.22 & 0.02 \\
       & -0.90 & 0.44 & 0.02 \\
    & -0.39 & 0.70 & 0.05 \\
   (300, -60)  & -0.46 & 0.15 & 0.03 \\
       & -0.97 & 0.23 & 0.03 \\
    & -0.30 & 0.71 & 0.07 \\
   & \multicolumn{2}{c}{{\bf{Positive correlation}}} \\
   (0, 3) & 0.44 & 0.50 & 0.08 \\
      & 0.38 & 0.68 & 0.07 \\
   & 0.59 & 0.64 & 0.07 \\
   (349, 7) & -0.16 & 0.51 & 0.02 \\
       & 0.07 & 0.56 & 0.03 \\
    & 0.39 & 0.57 & 0.02 \\
   (250, 8) & 0.04 & 0.24 & 0.03 \\
       & 0.23 & 0.24 & 0.03 \\
    & 0.25 & 0.35 & 0.02\\
    & \multicolumn{2}{c}{{\bf{Typical signal-to-noise}}} \\
   (200, -5) & -0.07 & 0.19 & 0.03 \\
       & -0.24 & 0.24 & 0.04 \\
    & 0.11 & 0.23 & 0.03 \\
   (183, -12) & -0.03 & 0.22 & 0.03 \\
       & -0.24 & 0.30 & 0.05 \\
    & 0.31 & 0.30 & 0.03 \\
   (160, 2) & -0.48 & 0.18 & 0.04 \\
       & -0.50 & 0.26 & 0.03 \\
    & -0.18 & 0.25 & 0.02 \\
   & \multicolumn{2}{c}{{\bf{Negative correlation}}} \\
   (150, 25) & -0.83 & 0.19 & 0.03 \\
       & -0.87 & 0.24 & 0.04 \\
    & -0.53 & 0.37 & 0.04 \\
   (140, -15) & -0.87 & 0.22 & 0.03 \\
       & -0.82 & 0.30 & 0.05 \\
    & -0.48 & 0.27 & 0.03 \\
   (260, 24) & -0.80 & 0.18 & 0.04 \\
       & -0.95 & 0.26 & 0.03 \\
    & -0.40 & 0.52 & 0.03 \\
  \hline
   \end{tabular}
\end{table}

Table~\ref{table:2} states the properties of the four regions displayed in Fig.~\ref{fig:corrplot}: the central coordinates of the region, the correlation coefficient between temperature and spectral index and the standard deviations with the region of the temperature and spectral index. For each region the \texttt{premise} correlation coefficient and standard deviations are listed first, then the GNILC values, then the 2013 values. Several other regions, additional to those shown in Fig.~\ref{fig:corrplot}, also have their properties listed in order to build up a picture of the parameter correlations for various types of regions. It can be seen that in the low signal-to-noise regions the 2013 method generally displays the least negative correlations alongside the largest parameter standard deviations. While, in the typical signal-to-noise regions, all three methods show negligible correlations between temperature and spectral index and similar parameter standard deviations. For the high signal-to-noise regions where both positive and negative correlations are seen the parameter standard deviations are similar for all three methods but while all three methods show similar levels of positive correlations, the 2013 methods shows the least negative correlation coefficients for all three negative correlation regions.

\subsection{Optical depth at 353\,GHz}

The first column of Fig.~\ref{fig:alltau} shows the all-sky $\tau_{353}$ as calculated by \texttt{premise}, 2013 and GNILC. The second column shows the difference maps between the three $\tau_{353}$ estimates and the third column shows the percentage difference maps between the three $\tau_{353}$ estimates. The comparisons appear to reveal the 2013 estimate to be noisier as, at high Galactic latitudes, the \texttt{premise} and GNILC $\tau_{353}$ estimates are closer in value to each other than the 2013 estimate. We also see again, from the percentage difference maps involving \texttt{premise}, the thin strip within the Galactic plane where the abundance of point sources prevents \texttt{premise} from providing accurate temperature estimates.  

The stellar reddening, E(B-V), due to extinction by interstellar dust provides a valuable and independent check on thermal dust column density estimates. \citet{pr2} make use of the SDSS quasar catalogue to determine correlations between E(B-V) and $\tau_{353}$. Since that analysis, a map of interstellar reddening \citep{green} has been made available; we make use of it and refer to it from herein as the Green reddening map.  
   
   \begin{figure}
   \centering
            {\includegraphics[width=0.7\linewidth]{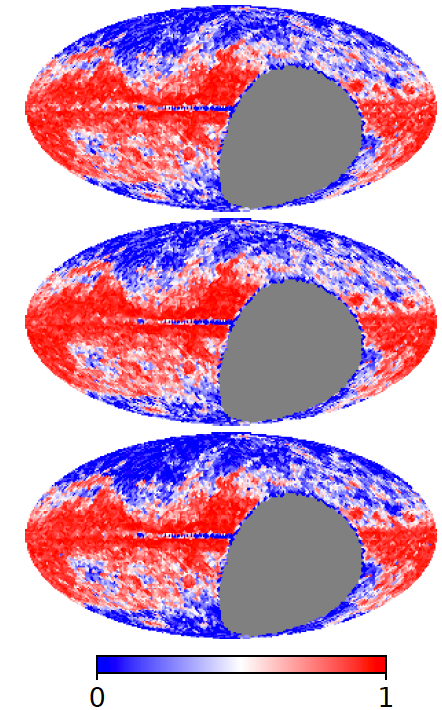}}
      \caption{E(B-V) and $\tau_{353}$ correlation maps for \texttt{premise} parameters {\it{(top}}), GNILC parameters {\it{(middle}}) and 2013 parameters ({\it{bottom}}). The colour scale represents the correlation coefficient values.}
         \label{fig:greenTau}
   \end{figure}
   
      \begin{figure}
   \centering
            {\includegraphics[width=0.99\linewidth]{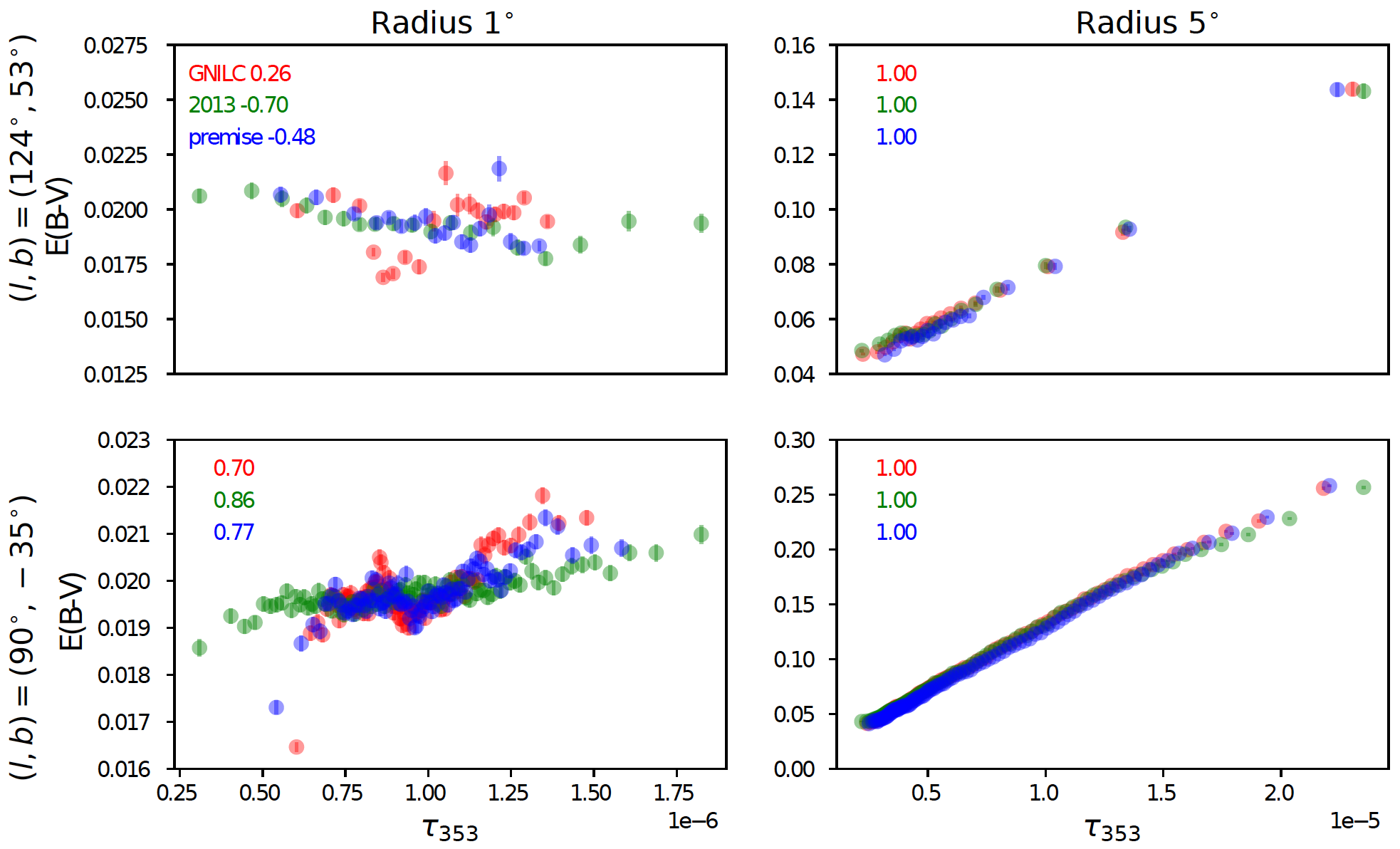}}
      \caption{E(B-V) versus $\tau_{353}$ scatter plots for two different regions and two different angular scales.}
         \label{fig:greenTauscat}
   \end{figure}
   
Fig.~\ref{fig:greenTau} presents correlation coefficient maps made by dividing the full sky into patches of 64X64 pixels, between $\tau_{353}$ and E(B-V) for optical depth values calculated by \texttt{premise}, GNILC and 2013. The grey patch within each map represents the 25 per cent of the sky absent within the Green reddening map. All three maps show strong positive correlations within high signal-to-noise regions. We select two circular regions for further investigation: a high signal-to-noise region centred at $(l,b) = (90^{\circ}, -35^{\circ})$ displaying strong positive correlations for all three estimates, and a low signal-to-noise region centred at $(l,b) = (124^{\circ}, 53^{\circ})$ displaying weak correlations. We consider radii of one and five degrees for each location. The relationships between $\tau_{353}$ and E(B-V) in these regions are shown in Fig.~\ref{fig:greenTauscat}. The data are binned into bins of 1000 data points for the five degree radius circles and 200 data points for the one degree radius circles. The data points for the one degree radius plots displays larger error bars as the error bars are the standard deviation within the bin divided by the square root of the bin number.  The correlation coefficients for the binned data are given on the plots.

From Fig.~\ref{fig:greenTauscat} it can be seen that at large angular scales the correlations between optical depth estimates and interstellar reddening are indistinguishably strong for all three optical depth estimates. At small angular scales however, the signal-to-noise ratio within the region in question has a notable effect on the correlation coefficient. For high signal-to-noise regions all three $\tau_{353}$ estimates show strong positive correlations but within the low signal-to-noise regions weak positive correlations and even negative correlations can be seen. The GNILC estimate is more tightly correlated with E(B-V) at small scales in high signal-to-noise regions, but there are regions in the sky where the premise or 2013 estimates follow more closely the reddening than GNILC. Regardless of which method is enlisted to deal with the contamination of CIBA there are regions at low angular resolution which remain affected by this contamination. To try and distinguish between the three methods we now extend our analysis to thermal dust radiance estimates.    

\subsection{Radiance maps from the MBB parameters}

            \begin{figure}
   \centering
            {\includegraphics[width=0.99\linewidth]{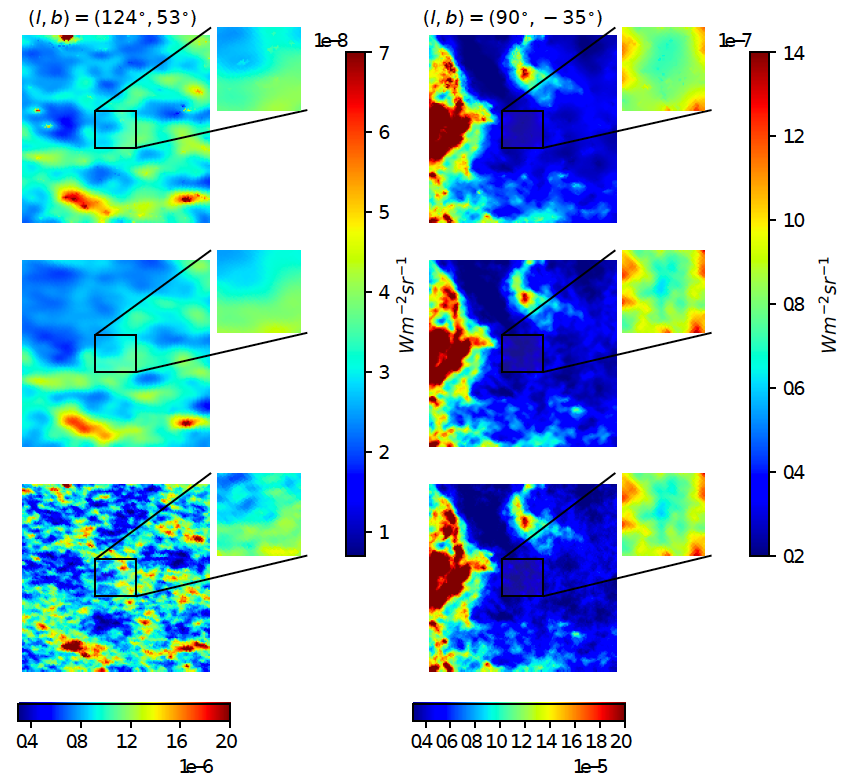}}
      \caption{$\tau_{353}$ maps of five-by-five degrees within two different regions for \texttt{premise}, GNILC and 2013 {\it{(from top to bottom)}}. The insets zoom into the central one-by-one degree regions of each map and display the equivalent radiance map for each method. The central coordinate of each map is given above each column.}
         \label{fig:rad}
   \end{figure}
   
            \begin{figure}
   \centering
            {\includegraphics[width=0.99\linewidth]{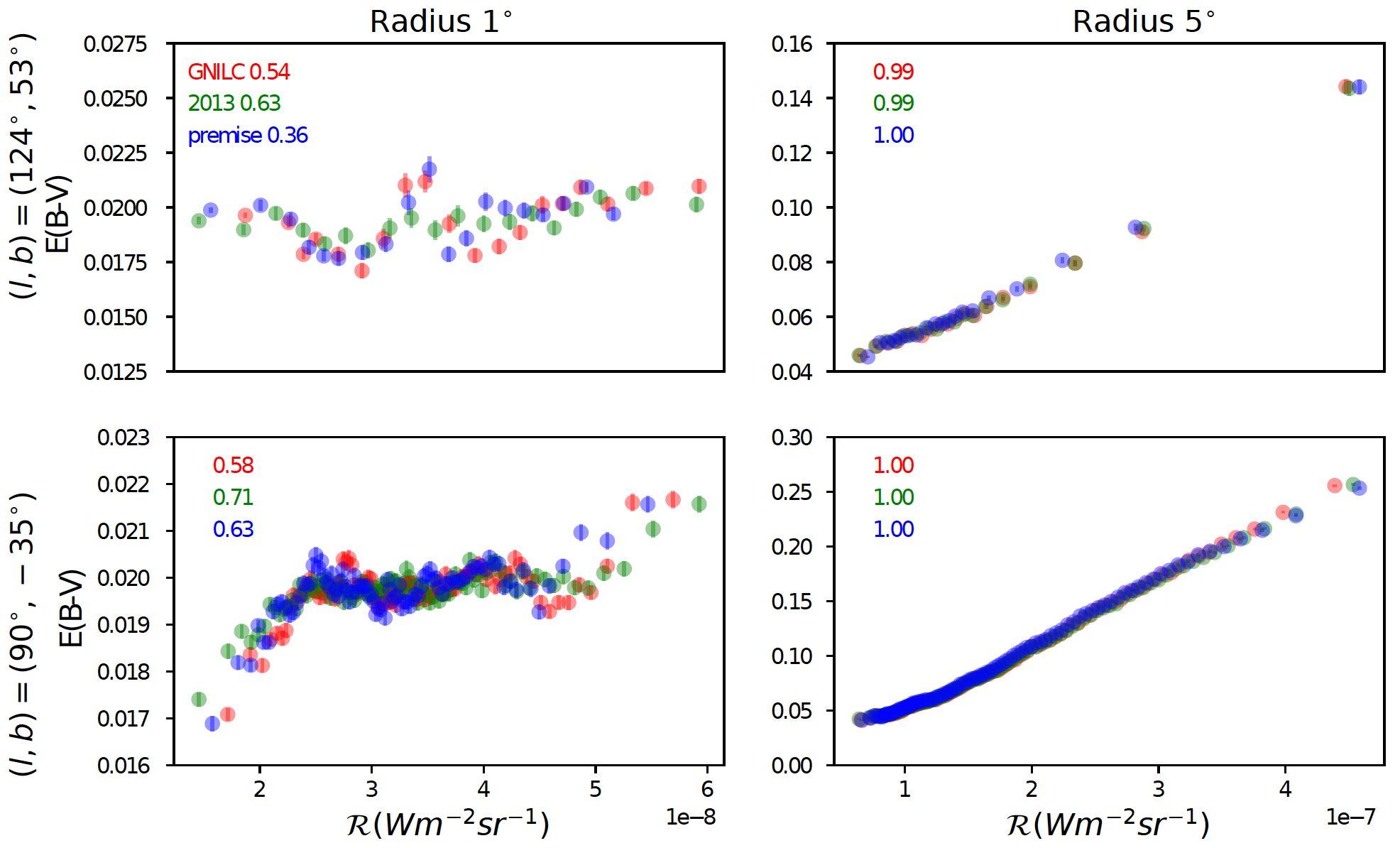}}
      \caption{E(B-V) versus radiance scatter plots for two different regions and two different angular scales.}
         \label{fig:greenRadscat}
   \end{figure}

\citep{pr2} have shown that thermal dust radiance, the integrated intensity over frequency, provides a cleaner way to view thermal dust estimates. This is because the CIBA decorrelate over frequency, leaving less of a residual impact in the radiance maps than within the optical depth estimates. We calculate radiance using the Gamma ($\Gamma$) and Riemann Zeta ($\zeta$) functions as in \citep{pr2}:
\begin{equation}
\mathcal{R} = \tau_{353} \frac{\sigma_{S}}{\pi} T^{4} \left (\frac{kT}{h\nu_{0}} \right)^{\beta} \frac{\Gamma (4 + \beta) \zeta (4 + \beta)}{\Gamma (4) \zeta (4)},
\end{equation} 
where $\sigma_{S}$ is the Stefan-Boltzmann constant, $h$ is the Planck constant, $k$ is the Boltzmann constant and $\nu_{0} = 3.53 \times 10^{11}$\,Hz. 

Fig.~\ref{fig:rad} shows both the optical depth at 353\,GHz and the radiances for \texttt{premise}, GNILC and 2013 within the low and high signal-to-noise regions of Fig.~\ref{fig:greenTauscat}. The five-by-five degree fields of view show the $\tau_{353}$ estimates while the zoomed in one-by-one degree insets show the radiance values. For the low signal-to-noise region ($(l,b) = (124^{\circ}, 53^{\circ})$) the GNILC $\tau_{353}$  estimate can be seen to display a higher degree of smoothing and so fewer spatial features than the 2013 and \texttt{premise} $\tau_{353}$ estimates. Whereas, within the high signal-to-noise region the three $\tau_{353}$ estimates look far more similar. The \texttt{premise} $\tau_{353}$ estimate has had thresholding applied to its wavelet coefficients and so appears smoother, however features at the 5 arcmin scale are clearly visible upon inspection.   

To fully interpret the radiance estimates shown in Fig.~\ref{fig:rad} we perform the same correlation coefficient analysis with E(B-V) as before, on the same two regions but this time for the radiance estimates. Fig.~\ref{fig:greenRadscat} reveals, as before, that the three methods are only distinguishable from each other across small angular scales. For the low signal-to-noise region (top left of Fig.~\ref{fig:greenRadscat}) there is an increase in the correlation coefficient for all three methods when compared to the $\tau_{353}$, one-by-one degree, low signal-to-noise scatter plot as the radiance calculation integrates the intensity over frequency and with increasing frequency the fractional percentage of CIBA in the total emission decreases. Fig.~\ref{fig:rad} shows the 2013 radiance estimates within the low signal-to-noise region to be higher than the other two estimates; this is due to the steep scaling of the radiance with temperature ($T^{4+\beta}$) and the fact that the 2013 mean temperature estimate within this one-by-one degree region is 22.0 K as opposed to the 20.1/20.2\,K mean temperature predicted by GNILC/ \texttt{premise}.

For the high signal-to-noise region (bottom left of Fig.~\ref{fig:greenRadscat}) the correlation coefficients decrease from optical depth to radiance. The signal-to-noise ratio is high enough within this region for the CIBA contamination within the optical depth estimates to no longer be the dominant factor. The $\tau_{353}$ maps in Fig.~\ref{fig:rad} show all three estimates to have similar structures and resolutions for the highest values. Within this high signal-to-noise region it is the variations in MBB temperature and spectral index which are responsible for the decreases in correlation coefficient. Fig.~\ref{fig:1deg} shows the temperature and spectral index estimates for each of the three methods per pixel of the one degree radius circle centred at $(l,b) = (90^{\circ}, -35^{\circ})$. The GNILC temperature and spectral indices display the largest range of spectral indices and temperatures as well as a strong anti-correlation between these two parameters. This may be an example of the GNILC filtering smoothing out actual thermal dust emission as opposed to just the CIBA. The one-sigma deviation in temperature and spectral index for the 2013 estimates within the region shown in Fig.~\ref{fig:1deg} are larger than those for GNILC and \texttt{premise}. 

           \begin{figure}
   \centering
            {\includegraphics[width=0.7\linewidth]{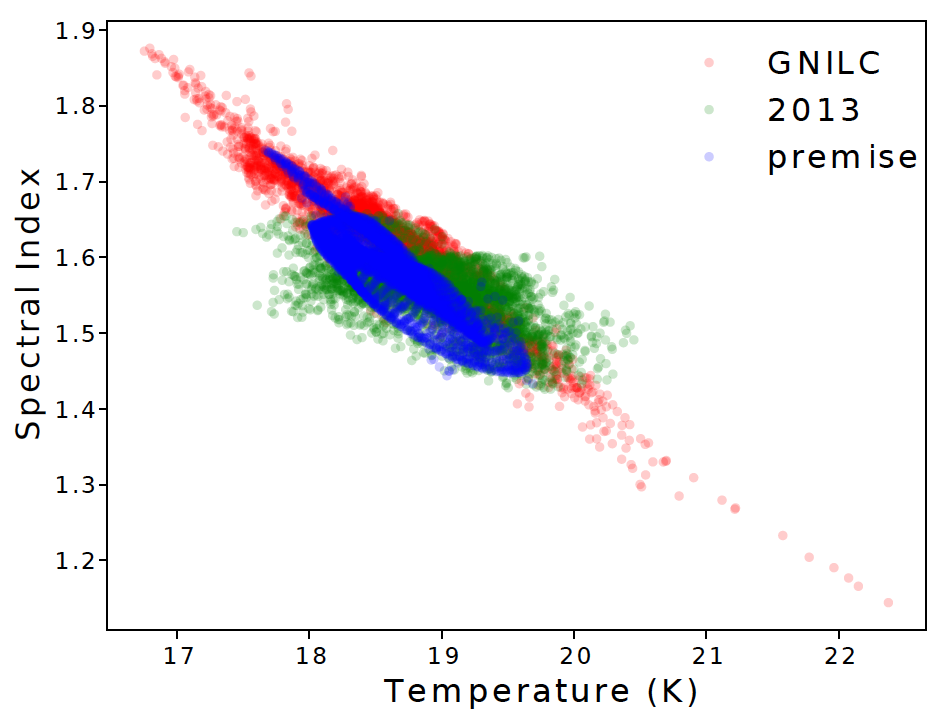}}
      \caption{Temperature versus spectral index scatter plot for the pixels contained within the one degree radius circle centred at $(l,b) = (90^{\circ}, -35^{\circ})$. }
         \label{fig:1deg}
   \end{figure}

\subsection{MBB parameters in detail}
\label{sec:clouds}

For a closer look at the differences between each of the MBB parameter estimates Fig.~\ref{fig:cloudsMBB} presents the $T$, $\beta$ and $\tau_{353}$ maps for four well-known molecular clouds: Spider, Draco, Taurus and Orion. Spider and Draco are lower in intensity than Taurus and Orion and so their $\tau_{353}$ maps are presented using a different colour scale. For the two high-intensity clouds it is hard to determine by eye significant differences between the three $\tau_{353}$ estimates, however for Spider and particularly Draco the 2013 $\tau_{353}$ maps appear to be slightly noisier. The GNILC $\beta$ values for Draco and Spider are generally higher than the 2013 and \texttt{premise} estimates. For the Spider region the 2013 temperature estimates appear to be anti-correlated with the 2013 $\tau_{353}$ estimate; this may be indicative of fitting an MBB in the presence of non-negligible noise. Within Draco the GNILC $T$ and $\beta$ estimates also appear to display correlations with their $\tau_{353}$ map. 

               \begin{figure*}
         \centering
            {\includegraphics[width=0.39\linewidth]{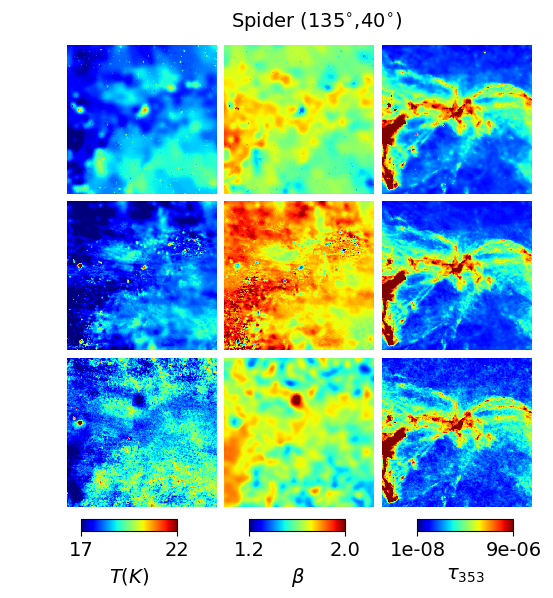}}
            {\includegraphics[width=0.39\linewidth]{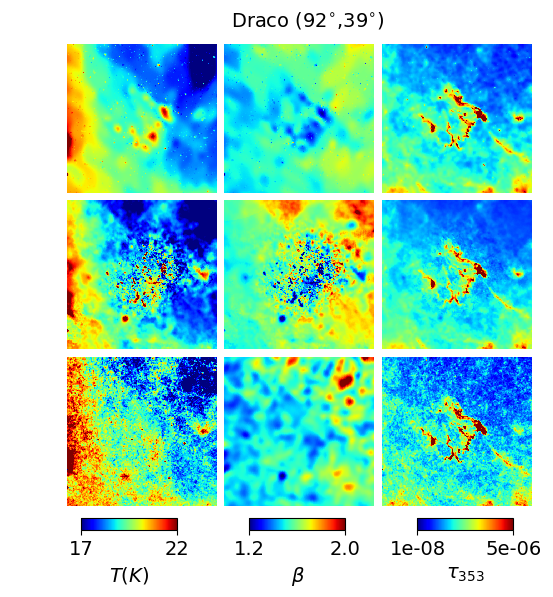}} \\
            {\includegraphics[width=0.39\linewidth]{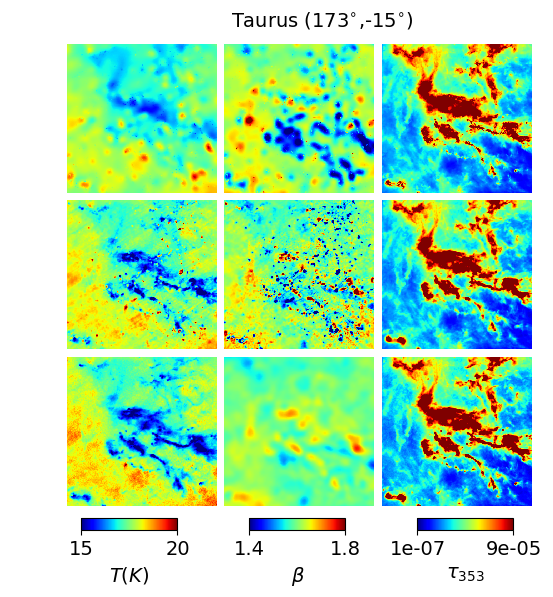}}
            {\includegraphics[width=0.39\linewidth]{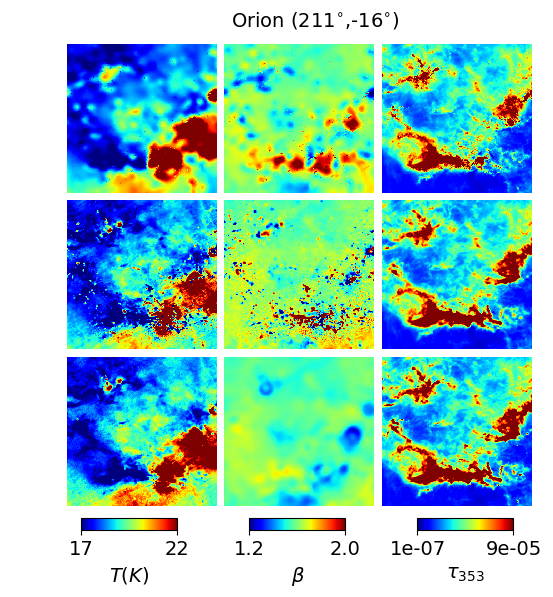}} 
      \caption{The MBB parameters for four molecular clouds as calculated by \texttt{premise} {\it{(top)}} , GNILC {\it{(middle)}} and 2013 {\it{(bottom)}}. The images are all $12^{\circ}$ by $12^{\circ}$.}
         \label{fig:cloudsMBB}
   \end{figure*}
   
               \begin{figure}
         \centering
            {\includegraphics[width=0.32\linewidth]{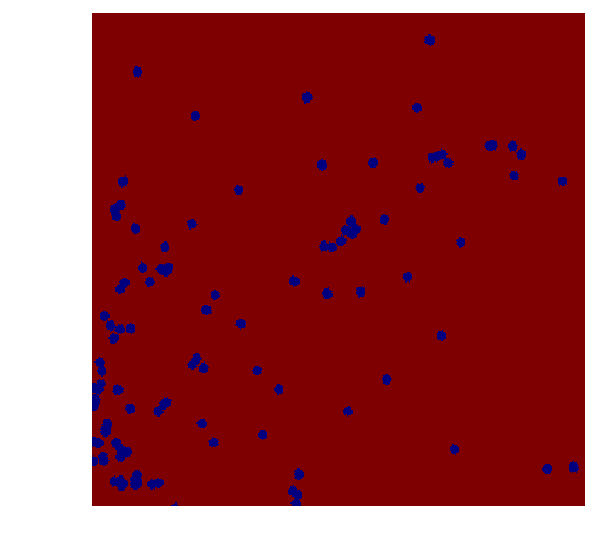}}
            {\includegraphics[width=0.32\linewidth]{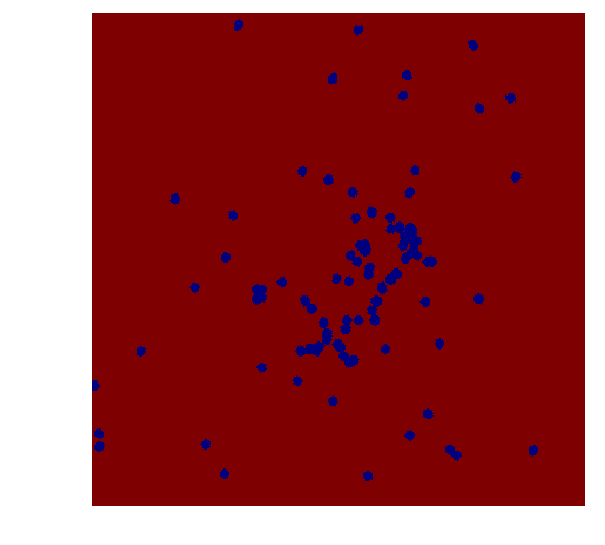}} \\
            {\includegraphics[width=0.32\linewidth]{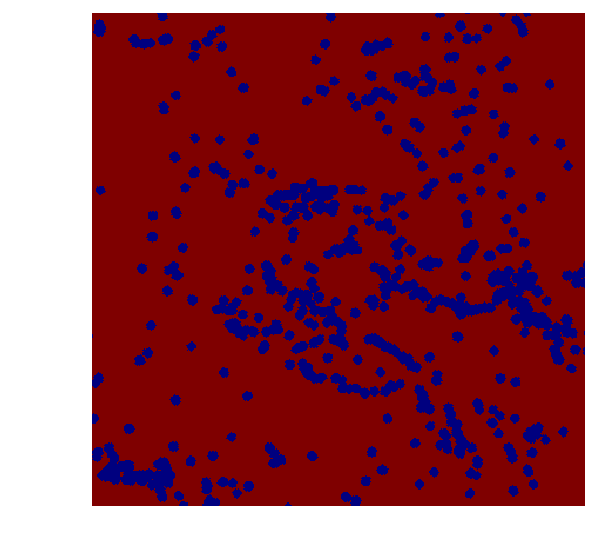}}
            {\includegraphics[width=0.32\linewidth]{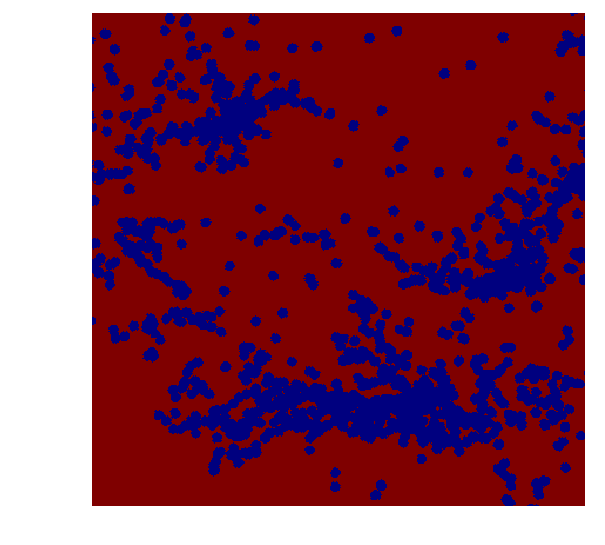}} 
      \caption{The combined {\it{Planck}} HFI point source mask for the Spider {\it{(top left)}} , Draco {\it{(top right)}} , Taurus {\it{(bottom left)}}  and Orion {\it{(bottom right)}}  molecular clouds.}
         \label{fig:cloudsMask}
   \end{figure}

For the two high-intensity molecular clouds the only notable differences between the three temperature estimates are the visible effects of the wavelet thresholding applied to the \texttt{premise} estimates and the additional point-like features present only in the GNILC spectral index and temperature estimates. As the 2013 $\beta$ estimates are smoothed to 30 armcin they show less features than the GNILC and \texttt{premise} estimates.

Overall for the two low-intensity molecular clouds \texttt{premise} and GNILC show more agreement in temperature with each other as 2013 seems noisier, while \texttt{premise} and 2013 show more agreement in $\beta$ with each other than GNILC. This may be due to the GNILC filtering which imposes higher degrees of smoothing for lower signal-to-noise regions because for the high-intensity molecular clouds, where the GNILC resolution will be close to 5 arcmin, the \texttt{premise} and GNILC $\beta$ estimates are far more similar to each other. 

In Draco, Taurus and Orion the GNILC $\beta$ maps display some point-like features. All three methods deal with point sources differently, the 2013 methodology leaves point sources in the total flux maps while \texttt{premise} and GNILC mask out point sources. \texttt{premise} uses the {\it{Planck}} HFI point source masks and sets the MBB parameter values for pixels within these masks to the average value of their neighbouring pixels (the neighbourhood being determined by the super-pixel areas). GNILC creates its own point source masks from the Planck Catalogue of Compact Sources and inpaints the total flux maps within these masks. We display the combined {\it{Planck}} HFI point source mask (the 353, 545 and 857\,GHz point sources masks multiplied together) in Fig.~\ref{fig:cloudsMask} to demonstrate the correlations between the location of point sources and the point-like features in the GNILC $\beta$ maps. However \texttt{premise} also sees different spectral indices with the Draco cloud centre, though not in the form of compact clusters of low $\beta$ values, so this may be indicative of the MBB model not providing the best fit within molecular clouds. There are indications from both \texttt{premise} and GNILC, not only within Draco but also the high-intensity molecular clouds of Taurus and Orion, that a spectral change accompanies, or generates, the absolute rise in opacity when going from the diffuse atomic to dense molecular media.

\subsection{Thermal dust maps}
\label{sec:dustmaps}

         \begin{figure*}
   \centering
            {\includegraphics[width=0.9\linewidth]{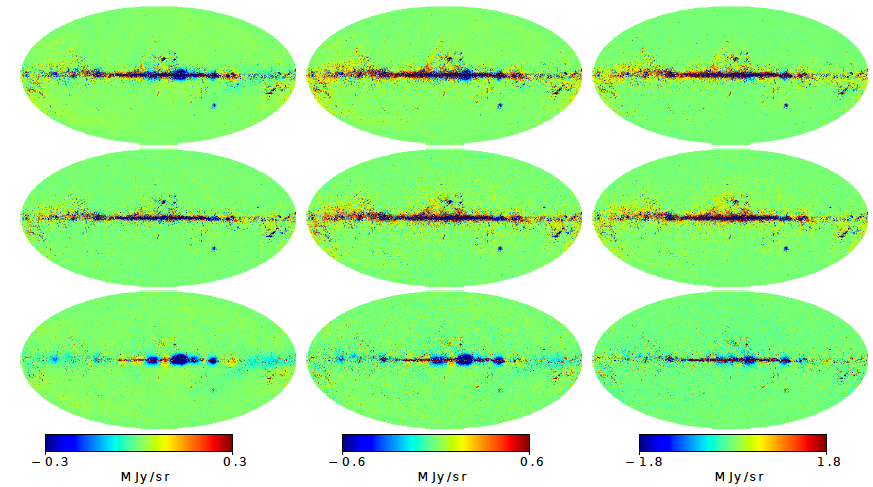}}
      \caption{Difference maps between the two thermal dust intensity estimates at each frequency: 353, 545 and 857\,GHz {\it{(left to right)}}. {\it{(Top:)}} $\rm{premise_{MBB}}$ - $\rm{GNILC_{MBB}}$, {\it{(middle:)}} $\rm{premise_{MBB}}$ - $\rm{2013_{MBB}}$, {\it{(bottom:)}} $\rm{2013_{MBB}}$ - $\rm{GNILC_{MBB}}$. }
         \label{fig:diffMapsMBB}
   \end{figure*}

           \begin{figure}
         \centering
            {\includegraphics[width=0.42\linewidth]{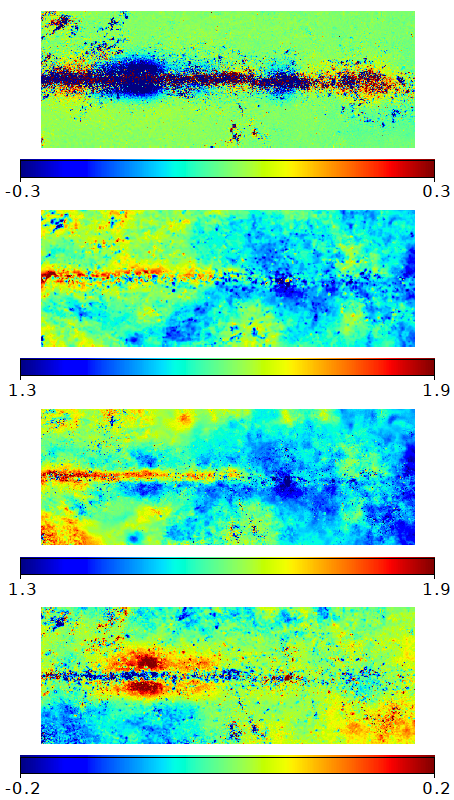}}
           {\includegraphics[width=0.42\linewidth]{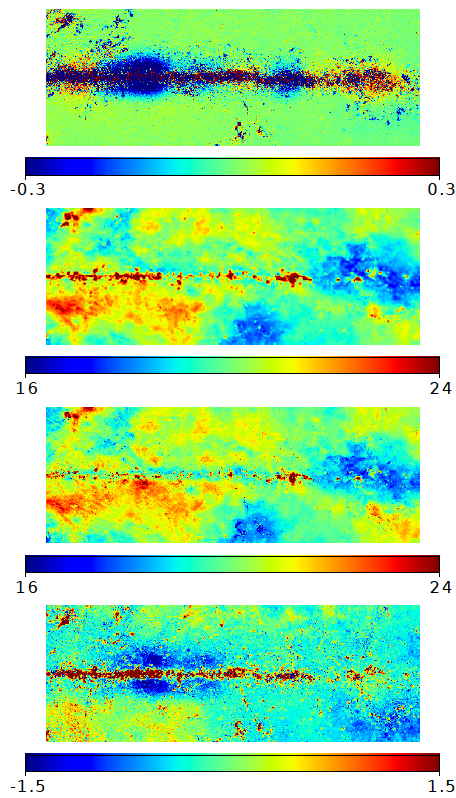}}
      \caption{{\it{Left:}} Cartesian projection of the central Galactic plane region of the {\it{Top:}} \texttt{premise} 353\,GHz dust estimate minus GNILC 353\,GHz dust estimate, {\it{upper middle:}} \texttt{premise} spectral index map, {\it{lower middle:}} GNILC spectral index map, {\it{bottom:}} (\texttt{premise} - GNILC) spectral index map. {\it{Right:}} Same as on the left but for temperature.}
         \label{fig:spots}
   \end{figure}
   
           \begin{figure*}
   \centering
            {\includegraphics[width=0.9\linewidth]{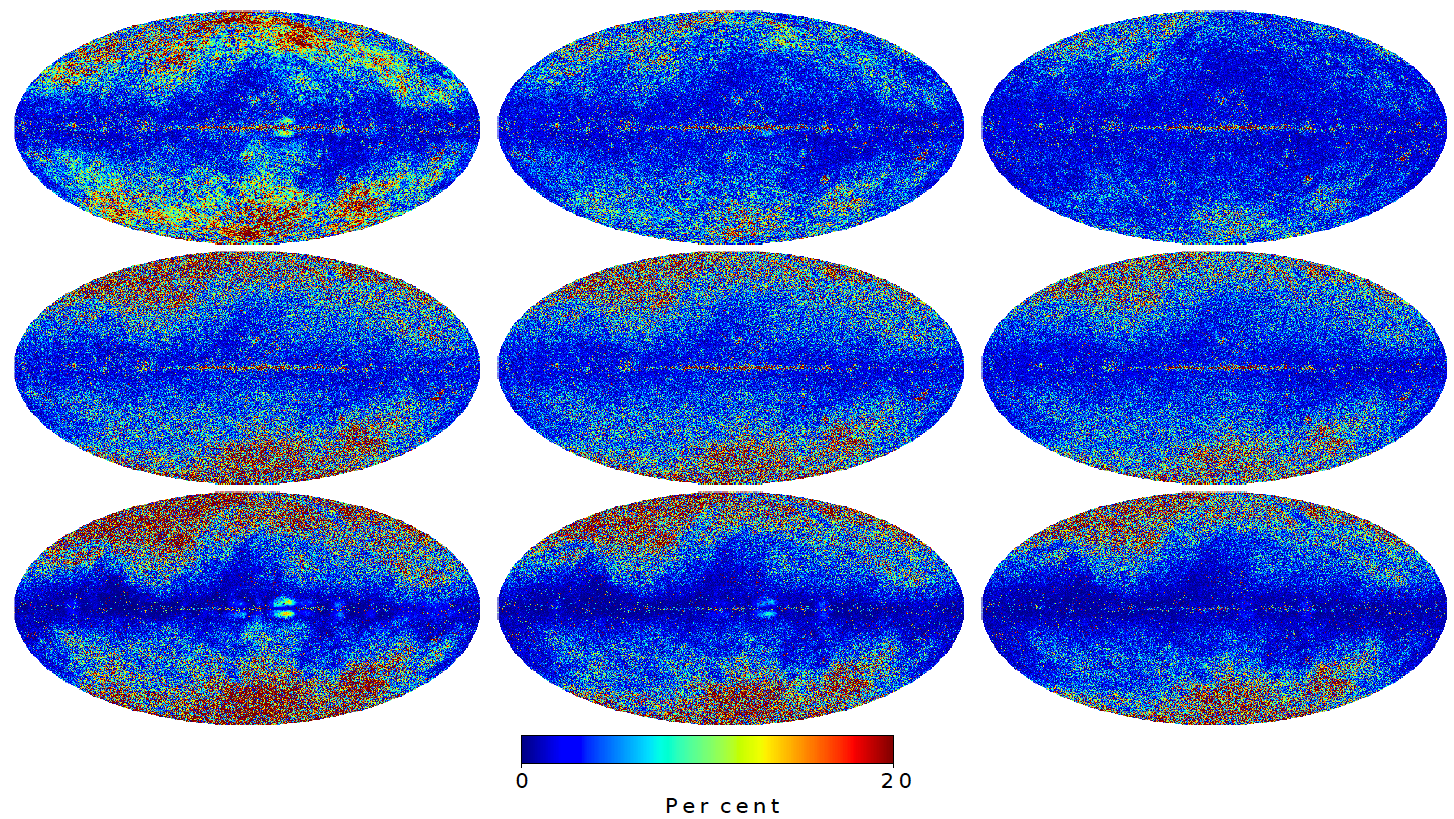}}
      \caption{Percentage difference maps between the MBB thermal dust intensity estimates at each frequency: 353, 545 and 857\,GHz {\it{(left to right)}}.{\it{(Top:)}} $\rm{premise_{MBB}}$ - $\rm{GNILC_{MBB}}$, {\it{(middle:)}} $\rm{premise_{MBB}}$ - $\rm{2013_{MBB}}$, {\it{(bottom:)}} $\rm{2013_{MBB}}$ - $\rm{GNILC_{MBB}}$. }
         \label{fig:perMapsMBB}
   \end{figure*}
   
           \begin{figure*}
   \centering
            {\includegraphics[width=0.9\linewidth]{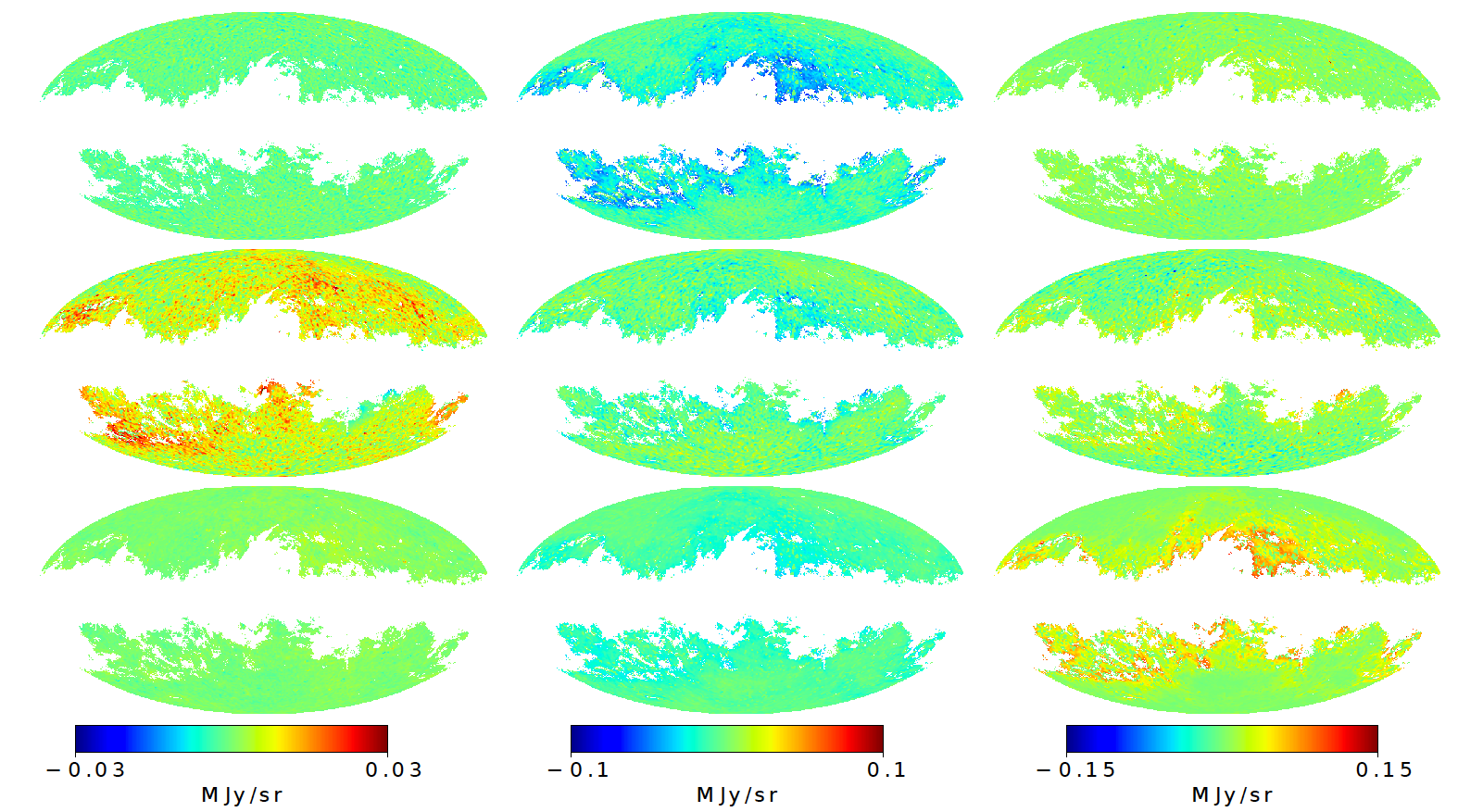}}
      \caption{CIBA plus instrumental noise maps at $1^{\circ}$ resolution made by ({\it{Top:}}) $\rm{premise_{MBB}}$,  ({\it{middle:}}) $\rm{GNILC_{MBB}}$ and ({\it{bottom:}}) $\rm{2013_{MBB}}$ at 353, 545 and 857\,GHz ({\it{left to right)}}.}
         \label{fig:mbbCIBsmooth}
   \end{figure*}

In this section we compare the \texttt{premise} thermal dust intensity estimates to those produced by GNILC and 2013. These dust maps have been produced by reconstructing the MBB parameters:
\begin{equation}
 I_{\rm{MBB}} = \tau_{353} \times B(T, \nu) \times \left(\frac{v}{353}\right)^{\beta} \times 10^{20} \times CC^{-1} {\rm{(MJy/sr)}},
\end{equation}

 Note: we need to multiply the MBB models by $10^{20}$ for the conversion to MJy/sr and divide them by their own colour corrections (formed using each method's temperature and spectral index estimates). We re-apply the effects of the passbands so that these total intensity estimates may be visually compared with the observation data ({\it{Planck}} frequency maps), which have not been colour corrected. 

Fig.~\ref{fig:diffMapsMBB} shows the difference between the $\rm{premise_{MBB}}$ and $\rm{GNILC_{MBB}}$ /$\rm{2013_{MBB}}$  intensity estimates at each frequency; as the Galactic plane has the highest thermal dust signal this region dominates these maps. At 353\,GHz globules of negative differences can be seen in the $\rm{premise_{MBB}}$/$\rm{GNILC_{MBB}}$ and $\rm{2013_{MBB}}$/$\rm{GNILC_{MBB}}$ difference maps along the Galactic plane, this is also the case at 545\,GHz and again but too a lesser degree at 857\,GHz. The most prominent `globule' at all frequencies is situated just to the right of the Galactic centre. Fig.~\ref{fig:spots} shows this particular region in detail within the \texttt{premise} 353\,GHz dust estimate minus GNILC 353\,GHz dust estimate, the GNILC temperature and spectral index maps, the \texttt{premise} temperature and spectral index maps and the temperature and spectral index difference maps. The globular feature is also present in the temperature and spectral index difference maps and is caused by GNILC discerning temperature and spectral index features around the Galactic plane that are not seen by \texttt{premise}. Within and around the Galactic plane resolution is not an issue as both the \texttt{premise} and GNILC results are presented at 5 arcmin here. It is possible that these globular features along the Galactic plane relate to actual changes in the dust MBB temperature and spectral index but as neither \texttt{premise} nor 2013 see such features they could also be a feature of the GNILC processing. 
   
Fig.~\ref{fig:perMapsMBB} shows percentage difference maps at each frequency: the absolute difference between the $\rm{premise_{MBB}}$ and $\rm{GNILC_{MBB}}$ /$\rm{2013_{MBB}}$ dust estimates divided by the $\rm{GNILC_{MBB}}$ /$\rm{2013_{MBB}}$ dust map at each frequency. For all frequencies the largest differences are at high latitudes where the dust signal-to-noise is weakest and where the GNILC algorithm applies the greatest degree of smoothing. Along $b=0$ the $\rm{premise_{MBB}}$ thermal dust estimate differs from the $\rm{2013_{MBB}}$  and $\rm{GNILC_{MBB}}$ to a larger extent than the $\rm{2013_{MBB}}$ and $\rm{GNILC_{MBB}}$ models differ. This highlights the difficulties the {\texttt{premise}} algorithm has in recovering model parameters across heavily masked regions. The GNILC globules are again visible across the Galactic plane at 353\,GHz for both the $\rm{premise_{MBB}}$/$\rm{GNILC_{MBB}}$ and $\rm{2013_{MBB}}$$\rm{GNILC_{MBB}}$ percentage difference maps. 

      \begin{figure}
   \centering
            {\includegraphics[width=0.9\linewidth]{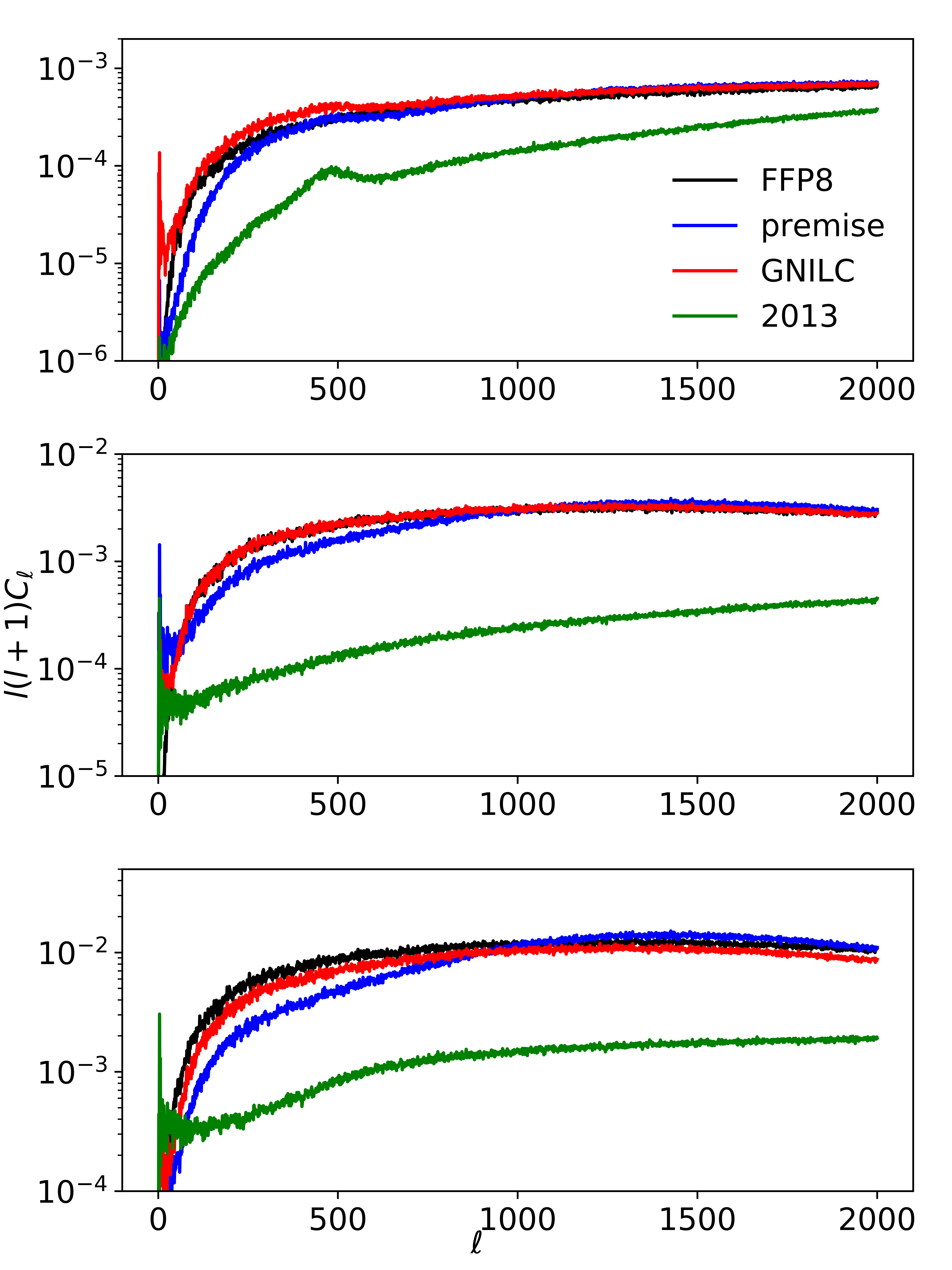}}\\
      \caption{Power spectra of the CIBA plus instrumental noise estimates alongside the FFP8 simulations at 353, 545 and 857\,GHz ({\it{top to bottom)}}.}
         \label{fig:mbbCIBvarmaps}
   \end{figure}
   
\subsection{CIB maps}

Subtracting the MBB intensity estimates (calculated in the previous section) from the total intensity maps (after the G-MCA estimate of the CMB has been removed), then subtracting of the large-scale CIB offsets results in maps of the CIBA plus instrumental noise, as shown in Fig.~\ref{fig:mbbCIBsmooth}. These CIBA plus instrumental noise maps are smoothed to one degree FWHM. The expectation is for these maps to resemble Gaussian noise as the Poisson term of the CIBA only starts to dominate at angular scales smaller than the {\it{Planck}} HFI resolutions. We create a Galactic plane mask based on the {\texttt{premise}} $\tau_{353}$ estimate, choosing to mask out the regions where  $\tau_{353}$ is more than 1.3 times greater than the median value. This mask leaves 57 per cent of the sky still visible.  

At 353\,GHz both the $\rm{premise_{MBB}}$ and the $\rm{2013_{MBB}}$ residual maps are the cleanest while the $\rm{GNILC_{MBB}}$ residual map displays thermal dust contamination all over. At 545\,GHz the $\rm{premise_{MBB}}$ residual map shows negative intensity around the Galactic plane mask implying larger than true thermal dust estimates. At 857\,GHz the $\rm{2013_{MBB}}$ residual map shows the highest level of contamination, followed by $\rm{GNILC_{MBB}}$ then $\rm{premise_{MBB}}$. The $\rm{premise_{MBB}}$ residual map is cleaner than the $\rm{GNILC_{MBB}}$ residual at 353\,GHz and 857\,GHz and worse closer to the Galactic plane at 545\,GHz.
   
The FFP8 CIBA plus instrumental noise simulations provide estimates against which the $\rm{premise_{MBB}}$, $\rm{GNILC_{MBB}}$ and $\rm{2013_{MBB}}$ residual maps can be compared. In Fig.~\ref{fig:mbbCIBvarmaps} we present the various CIBA plus instrumental noise estimates as power spectra. The power spectra are formed from the full resolution residual maps. We use our Galactic plane mask to weight the $a_{l,m}$ to take account of the fact that these are partial sky power spectra. At 353\,GHz, if the FFP8 simulations are taken to represent the true CIBA plus instrumental noise, then the $\rm{GNILC_{MBB}}$ residual appears to overestimate the level of the CIBA plus instrumental noise, implying an underestimation in their thermal dust estimate at 353\;GHz. At small angular scales, the FFP8 simulations, $\rm{premise_{MBB}}$ and $\rm{GNILC_{MBB}}$ residuals estimates of CIBA plus instrumental noise are all in agreement. At large angular scales, however, we see a consistent under-prediction of CIBA plus instrumental noise for $\rm{premise_{MBB}}$ and $\rm{2013_{MBB}}$ residuals. 

            \begin{figure}
         \centering
            {\includegraphics[width=0.99\linewidth]{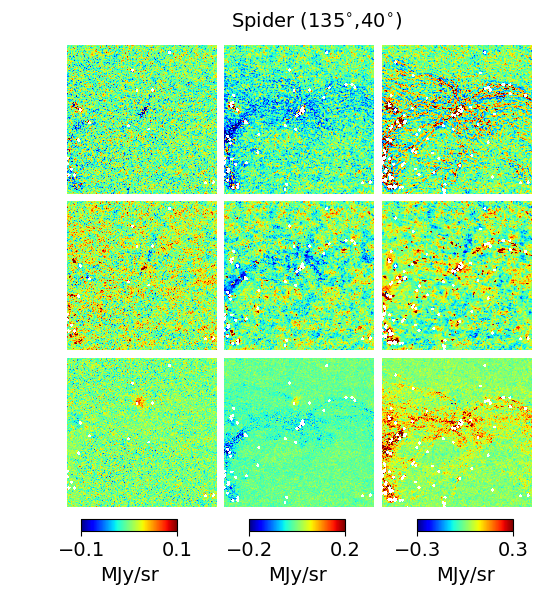}}\\
            {\includegraphics[width=0.99\linewidth]{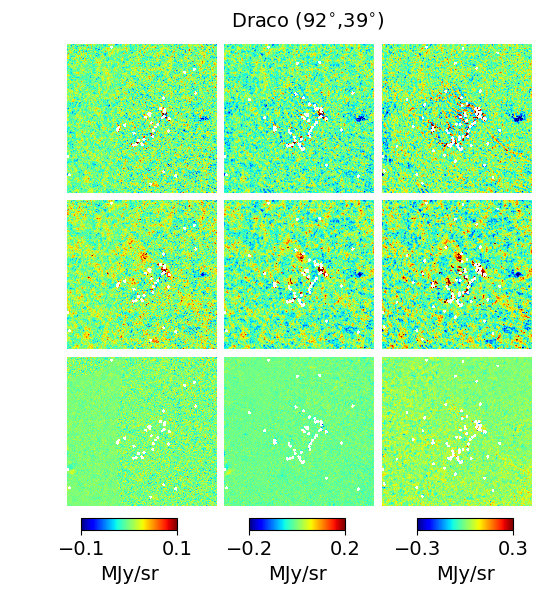}} 
      \caption{The CIBA plus instrumental noise estimates at 353, 545 and 857\,GHz  {\it{(from left to right)}} for two molecular clouds as calculated by \texttt{premise} {\it{(top)}} , GNILC {\it{(middle)}} and 2013 {\it{(bottom)}}. Point sources within have been masked out using the {\it{Planck}} HFI masks.}
         \label{fig:cloudsCIB}
   \end{figure}

For an in-depth look at the CIBA plus instrumental noise estimates we consider two of the four molecular clouds shown in section~\ref{sec:clouds}: Draco and Spider. The Taurus and Orion molecular cloud are within our Galactic plane mask and so do not feature in our CIB analysis. Fig.~\ref{fig:cloudsCIB} shows the CIBA plus instrumental noise estimates for the three methods at frequencies 353, 545 and 857\,GHz within Spider and Draco. We have applied the {\it{Planck}} HFI masks for each frequency as the total intensity data are used to form the CIBA plus noise estimates. Within both the Spider and Draco molecular clouds, the $\rm{GNILC_{MBB}}$ residual has more regions of high intensity at 353\,GHz than the other two residuals. For the Spider molecular cloud both the $\rm{premise_{MBB}}$ and $\rm{2013_{MBB}}$ residuals show a strong under-prediction of the CIBA plus instrumental noise at 545 and 857\,GHz as the outline of the Spider molecular cloud is still present. Whereas, for the Draco molecular cloud no distinct features are seen; yet the $\rm{2013_{MBB}}$ residuals at 353, 545 and 857\,GHz display a lower overall variance than the other two residuals.   


\section{Conclusions}

We have presented the first all-sky, 5 arcmin resolution, estimates of thermal dust MBB parameters formed from the {\it{Planck}} PR2 353, 545 and 857\,GHz maps as well the IRAS 3000\,GHz data. Our focus has been on obtaining as accurate model parameters as possible and to gauge our success we have made comparisons with two other methods which produce MBB parameter maps: GNILC and the 2013 two-step pixel-by-pixel fit of \citet{pr2} updated to use PR2 data. The main contaminant to obtaining pure thermal dust maps is the presence of the CIBA, this is typically dealt with by using some degree of smoothing. We believe that the use of smoothing increases the risk of averaging out thermal dust signal itself as well as the CIBA and seek to avoid this technique by instead exploiting the sparse nature of astrophysical foregrounds within the wavelet domain.

The work presented in this paper is the application of our algorithm \texttt{premise}, as presented in \citet{sparse}, to {\it{Planck}} HFI data. Our method has three main steps: filtering of the total intensity maps, an MBB fit to adaptive super-pixel areas and then a refinement step to provide full resolution parameter maps. Our filtering follows the philosophy of the GNILC filtering: we use noise plus CIBA estimates to identify noise dominated areas within the total intensity maps at each wavelet scale, we also bin our data thus performing a similar averaging-out to smoothing. However, we recapture intensity variations at the 5 arcmin scale by adding a sparse penalisation factor to the least squares optimisation. We also use a less stringent definition than GNILC for what constitutes a noise-dominated area. As a result we ensure that no thermal dust signal is lost within the filtering, the price of this being that a non-negligible level of CIBA makes it through our filtering process. The filtered maps are then divided up into super-pixels, the areas of which are chosen by the algorithm specially to provide the smallest pixel area within regions where the MBB parameters vary the most. The MBB fit to super-pixels helps to average out the remaining CIBA within our filtered maps and provides accurate initial guesses at the true parameter values. Lastly, we return to the total intensity data to perform a gradient descent at each full resolution pixel, using our initial, accurate MBB parameter guesses to seed the descent. Any thermal dust signal smoothed out through the super-pixel averaging is recaptured as well as, again, some low-level interference from the CIBA; therefore we transform the maps into the wavelet domain one last time to remove the wavelet coefficients (at each wavelet scale) where the signal level is below that of the noise.

We find the mean MBB temperature and spectral index across the full sky to be 19.5\,K and 1.54. Our temperature mean is in strong agreement with that of GNILC (19.4\,K) and 2013 (19.8\,K), while our mean spectral index is slightly lower than GNILC (1.60) but very similar to 2013 (1.53). Specific regions of large spectral index values seem to be present only within the GNILC spectral index maps and the GNILC temperature and spectral index values display the strongest degree of anti-correlation within low signal-to-noise regions.

For optical depth at 353\,GHz the GNILC and \texttt{premise} estimates are in stronger agreement with each other than with the 2013 estimate at high Galactic latitudes. As the CIBA plus instrumental noise maps reveal the 2013 dust estimate to contain the largest CIBA contamination, the 2013 optical depth estimate is most likely to just be noisier as opposed to biased. All three estimates show strong correlation with E(B-V) measurements at angular scales of 5 degrees but we can not comment on the correlations at the 1 degree level as our analysis does not have the sensitivity to yield conclusive results at this angular scale. We reconstitute the MBB parameters for each method to produce thermal dust intensity estimates. It should be noted that GNILC provide their filtered dust maps for public use, it is not these estimates which we use in our analysis. Our interest lies in accurate representation of model parameters and so we reconstitute the MBB parameter maps for each method in order to provide additional tests on the parameters themselves. The difference maps between the dust estimates of all three methods reveal globular features within the 353 and 545\,GHz GNILC maps, which we believe are due to the GNILC filtering technique as opposed to actual astrophysical features. Within low signal-to-noise regions all the estimates disagree with each other at 353, 545 and 857\,GHz, though the \texttt{premise} and GNILC dust estimates display the largest percentage disagreement at 353\,GHz at high Galactic latitudes and the strongest agreement at 545 and 857\,GHz.

It is also seen that within the Galactic plane \texttt{premise} struggles to produce accurate parameter values due to the large number of point sources which obscure the thermal dust information. As the \texttt{premise} parameter maps have been formed from {\it{Planck}} HFI data after the {\it{Planck}} HFI point source masks have been applied, our results should be considered in conjunction with these masks. The \texttt{premise} parameter maps cannot be used to extract thermal dust information within the regions masked out by the {\it{Planck}} HFI point source masks. 

The CIBA plus instrumental noise maps appear to show thermal dust contamination at 353\,GHz within the GNILC residual, this may account for the percentage difference between the \texttt{premise}/2013 and GNILC thermal dust estimates at 353\,GHz at high latitudes. The power spectra reveal that there is a non-negligible level of CIBA remaining in the 2013 and \texttt{premise} thermal dust estimates, at all angular scales for 2013 and angular scales larger than $\sim$ 10 arcmin for \texttt{premise}. We suggest that our MBB parameter estimates are an improvement on those of 2013 as our method removes more of the CIBA and an improvement on those of GNILC as our method does not go as far as to remove some of the thermal dust signal itself.

\begin{acknowledgements}
This work is supported by the European Community through the grant LENA (ERC StG no. 678282) within the H2020 Framework Program. This research has made use of the SVO Filter Profile Service (http://svo2.cab.inta-csic.es/theory/fps/) supported from the Spanish MINECO through grant AyA2014-55216.
\end{acknowledgements}

\bibliographystyle{aa} 
 \bibliography{refs} 



\newpage

\appendix

\section{Super-pixel decomposition}\label{sec:apB}\label{sec:apC}

In the \texttt{premise} estimation procedure, the parameter initialisation step makes use of a decomposition of spherical maps into super-pixels. This decomposition requires to extract the neighbours of each pixel of {\texttt{HEALPix}} maps. Extracting these pixels can be done by projecting each neighbour onto a Euclidean grid. However, this would highly increase the computational cost of the initialisation step. It is actually far more convenient and computational more efficient to make profit of the nested hierarchical pixel ordering of {\texttt{HEALPix}} sampling scheme to extract super-pixels. This is eventually equivalent to a nearest-neighbour project of pixels' neighbours onto a Euclidean grid.\\
While this procedure leads to a fast processing of spherical maps, it however has a major drawback: super-pixel extraction is not perfect at the borders of so-called {\texttt{HEALPix}} faces ({\it i.e.} the largest super-pixels that can be extracted within the {\texttt{HEALPix}} hierarchical pixel ordering). As a consequence, when forming the super-pixel MBB fit on each 2D {\texttt{HEALPix}} face border effects are visible within the full-sky temperature and spectral index maps, not only at the level of the joins between the twelve 2D faces but also at the super-pixel boundaries. The majority of these border effects are erased by the parameter refinement step as the total flux maps do not contain such artefacts. However, within particularly low signal-to-noise regions the super-pixel border effects feature too prominently to be removed by wavelet thresholding therefore we invoke a method known as Cycle-spinning \citep{spin}. 

To cycle spin data one applies a rotational shift to the data, operates a function on the shifted data and then applies the inverse of the rotational shift thus changing the pixel location of any artefacts introduced to the solution by the function itself. If numerous spins are performed and their results averaged then the operational artefacts can be averaged out with any effect to the true solution. 

In the case of \texttt{premise} our initial temperature and spectral index estimates are formed from the full-sky intensity maps as follows:
\begin{itemize}
  \renewcommand{\labelitemi}{$\Rightarrow$}
 \item Full-sky maps  broken up into 12 faces
 \item Faces broken up into super-pixels
 \item A single MBB model is fit to each super-pix to determine $T$ and $\beta$
 \item $T$ and $\beta$ estimates per face are reconstructed to form all sky maps of $T$ and $\beta$
\end{itemize}
For cycle-spinning the above method is altered to: 
\begin{itemize}
  \renewcommand{\labelitemi}{$\Rightarrow$}
  \item Apply a rotational transform to the full-sky maps
 \item Full-sky maps  broken up into 12 faces
 \item Faces broken up into super-pixels
 \item An MBB model is fit to each super-pix to determine $T$ and $\beta$
 \item $T$ and $\beta$ estimates per face are reconstructed to form all sky maps of $T$ and $\beta$
 \item Apply the inverse of the rotational transform to the results
\end{itemize}

We apply six 2D rotational transforms to the full-sky maps, as described in Table ~\ref{tab:accstats}, and average the results of these six transforms plus the original (no transform) results together to form our initial estimates for temperature and spectral index.   

\begin{table}
 \caption{Two-dimensional rotations used to cycle spin the temperature and spectral index super-pixel maps.}
 \label{tab:accstats}
 \begin{tabular}{l| c| c| c| c| c| c| c}
{\bf{Axis}} & {\bf{ R0}} &{\bf{ R1}}  & {\bf{R2}}  & {\bf{R3}}  & {\bf{R4}}  & {\bf{R5}}  & {\bf{R6}} \\ 
\hline
X & 0$^{\circ}$ & 90$^{\circ}$ & 0$^{\circ}$ & -90$^{\circ}$ & 0$^{\circ}$ & 180$^{\circ}$ & 180$^{\circ}$\\
Y & 0$^{\circ}$ & 0$^{\circ}$ & 180$^{\circ}$ & 0$^{\circ}$ & -180$^{\circ}$ & 180$^{\circ}$ & -180$^{\circ}$\\
 \end{tabular}
\end{table}

 \end{document}